\documentclass[english,aps,superscriptaddress,showkeys,showpacs,prepri,twocolumn]{revtex4}
\usepackage[T1]{fontenc}
\pdfoutput=1
\usepackage[letterpaper]{geometry}
\usepackage{wrapfig,rotating}
\usepackage{units}
\usepackage{bbding}
\usepackage{amsmath}
\usepackage{amssymb}
\usepackage{graphicx}
\usepackage{esint}
\usepackage[colorinlistoftodos,prependcaption,textsize=tiny]{todonotes}

\usepackage[utf8]{inputenc}
\usepackage{lmodern} 

\makeatletter

\@ifundefined{textcolor}{}
{%
 \definecolor{BLACK}{gray}{0}
 \definecolor{WHITE}{gray}{1}
 \definecolor{RED}{rgb}{1,0,0}
 \definecolor{GREEN}{rgb}{0,1,0}
 \definecolor{BLUE}{rgb}{0,0,1}
 \definecolor{CYAN}{cmyk}{1,0,0,0}
 \definecolor{MAGENTA}{cmyk}{0,1,0,0}
 \definecolor{YELLOW}{cmyk}{0,0,1,0}
}

\usepackage{doi}
\usepackage{hyperref}

\makeatother

\usepackage{babel}
\begin{document}

\title{MHD modeling of a DIII-D low-torque QH-mode discharge and comparison to observations}

\author{J. R. King}
\affiliation{Tech-X Corporation, 5621 Arapahoe Ave. Boulder, CO 80303, USA}

\author{S. E. Kruger}
\affiliation{Tech-X Corporation, 5621 Arapahoe Ave. Boulder, CO 80303, USA}

\author{K. H. Burrell}
\affiliation{General Atomics, PO Box 85608, San Diego, CA 92186–5608, USA}

\author{X. Chen}
\affiliation{General Atomics, PO Box 85608, San Diego, CA 92186–5608, USA}

\author{A.M. Garofalo}
\affiliation{General Atomics, PO Box 85608, San Diego, CA 92186–5608, USA}

\author{R. J. Groebner}
\affiliation{General Atomics, PO Box 85608, San Diego, CA 92186–5608, USA}

\author{K. E. J. Olofsson}
\affiliation{Oak Ridge Associated Universities, Oak Ridge, TN 37831, USA}

\author{A.Y. Pankin}
\altaffiliation{Presently at Lawrence Livermore National Laboratory}
\affiliation{Tech-X Corporation, 5621 Arapahoe Ave. Boulder, CO 80303, USA}

\author{P.B. Snyder}
\affiliation{General Atomics, PO Box 85608, San Diego, CA 92186–5608, USA}

\date{draft \today}
\begin{abstract} 
Extended-MHD modeling of DIII-D tokamak [J.~L.~Luxon, Nucl.~Fusion 42, 614
2002] QH-mode discharges with nonlinear NIMROD [C.~R.~Sovinec et al., JCP 195,
355 2004] simulations saturates into a turbulent state, but does not saturate
when the steady-state flow inferred from measurements is not included. This is
consistent with the experimental observations of the quiescent regime on
DIII-D.  The simulation with flow develops into a saturated turbulent state
where the $n_\phi$=1 and 2 toroidal modes become dominant through an inverse
cascade. Each mode in the range of $n_\phi$=1-5 is dominant at a different
time. Consistent with experimental observations during QH-mode, the simulated
state leads to large particle transport relative to the thermal transport.
Analysis shows that the amplitude and phase of the density and temperature
perturbations differ resulting in greater fluctuation-induced convective
particle transport relative to the convective thermal transport.  Comparison to
magnetic-coil measurements shows rotation frequencies differ between the
simulation and experiment which indicates that more sophisticated extended-MHD
two-fluid modeling is required.  
\end{abstract}

\keywords{extended-MHD modeling, quiescent H-mode, peeling-ballooning modes, nonlinear simulation, tokamak pedestal}

\pacs{52.30.Ex 52.35.Py, 52.55.Fa, 52.55.Tn, 52.65.Kj}

\maketitle



  \newcommand{\vect}[1]{ \mathbf{#1}}
  \newcommand{\defn}{ \equiv}

  \newcommand{\lp}{\left(}
  \newcommand{\rp}{\right)}
  \newcommand{\lb}{\left[}
  \newcommand{\rb}{\right]}
  \newcommand{\la}{\left<}
  \newcommand{\ra}{\right>}

  \newcommand{\vf}{ \vect{f}}

  \newcommand{\vx}{\vect{x}}
  \newcommand{\vq}{\vect{q}}
  \newcommand{\vB}{\vect{B}}
  \newcommand{\vJ}{\vect{J}}
  \newcommand{\vA}{\vect{A}}
  \newcommand{\vE}{\vect{E}}
  \newcommand{\vV}{\vect{V}}
  \newcommand{\vF}{ \vect{F} }	
  \newcommand{\vU}{ \vect{U} }	
  \newcommand{\ddp}{\grad \cdot \Pi}
  \newcommand{\specheat}{\gamma_h}

  \newcommand{\grad}{\vect{\nabla}}
  \newcommand{\curl}[1]{\grad \times #1 }
  \newcommand{\dive}[1]{\grad \cdot #1 }
  \newcommand{\vdg}{\left(\vV \cdot \grad \right)}
  \newcommand{\bdg}{\left(\vB \cdot \grad \right)}
  \newcommand{\divV}{\grad \cdot \vV_1}
  \newcommand{\divVp}{\left( \grad \cdot \vV \right)}

  \newcommand{\dt}[1]{\frac{\partial #1}{\partial t}}
  \newcommand{\Dt}[1]{\frac{d #1}{dt}}
  \newcommand{\dpsi}[1]{\frac{\partial #1}{\partial \psi}}
  \newcommand{\dpsisq}[1]{\frac{\partial^2 #1}{\partial \psi^2}}
  
  \newcommand{\jac}{{\mathcal{J}}}
  \newcommand{\jaci}{{\mathcal{J}}^{-1}}
  \newcommand{\Pp}{ P^\prime }				
  \newcommand{\Vp}{V^\prime}
  \newcommand{\Vpp}{V^{\prime\prime}}
  \newcommand{\Vpo}{ \frac{V^\prime}{4 \pi^2}}
  \newcommand{\norm}{ P^\prime }		
  \newcommand{\RR}{ \psi }			
  \newcommand{\vR}{ \grad \RR }		
  \newcommand{\C}{ C }				
  \newcommand{\vC}{ \vect{\C} }		
  \newcommand{\vK}{ \vect{K} }			
  \newcommand{\vRsq}{ \mid \grad \RR \mid^2 }
  \newcommand{\vCsq}{ \C^2 }
  \newcommand{\vKsq}{ K^2 }
  \newcommand{\vBsq}{ B^2 }
  \newcommand{\vrr}{\frac{ \vR}{\vRsq} }
  \newcommand{\vbb}{\frac{ \vB}{B^2} }
  \newcommand{\vcc}{\frac{ \vC}{\vCsq} }
  \newcommand{\vjj}{\frac{ \vJ}{J^2} }
  \newcommand{\vkk}{\frac{ \vK}{\vKsq} }

  \newcommand{\R}{ \psi }
  \newcommand{\T}{ \Theta }
  \newcommand{\Z}{ \zeta }
  \newcommand{\A}{ \alpha }
  \newcommand{\U}{ u }
  \newcommand{\ve}{ \vect{e} }
  \newcommand{\vur}{ \vect{e}^\rho }
  \newcommand{\vut}{ \vect{e}^\Theta }
  \newcommand{\vuz}{ \vect{e}^\zeta }
  \newcommand{\vlr}{ \vect{e}_\rho }
  \newcommand{\vlt}{ \vect{e}_\Theta }
  \newcommand{\vlz}{ \vect{e}_\zeta }
  \newcommand{\gr}{ \grad \R }
  \newcommand{\gt}{ \grad \Theta }
  \newcommand{\gz}{ \grad \zeta }
  \newcommand{\ga}{ \grad \alpha }
  \newcommand{\gu}{ \grad \U }
  \newcommand{\dr}[1]{ \frac{\partial #1}{\partial \R} }
  \newcommand{\dT}[1]{\frac{\partial #1}{\partial \Theta}}
  \newcommand{\dz}[1]{\frac{\partial #1}{\partial \zeta}}
  \newcommand{\dU}[1]{\frac{\partial #1}{\partial \U}}
  \newcommand{\drs}[1]{ \frac{\partial^2 #1}{\partial \R^2} }
  \newcommand{\dTs}[1]{\frac{\partial^2 #1}{\partial \Theta^2}}
  \newcommand{\drt}[1]{\frac{\partial^2 #1}{\partial \R \partial \Theta}}
  \newcommand{\dzs}[1]{\frac{\partial^2 #1}{\partial \zeta^2}}
  \newcommand{\grr}{ g^{\R \R} }
  \newcommand{\grt}{ g^{\R \Theta} }
  \newcommand{\grz}{ g^{\R \zeta} }
  \newcommand{\gtz}{ g^{\Theta \zeta} }
  \newcommand{\gtt}{ g^{\Theta \Theta} }
  \newcommand{\gzz}{ g^{\zeta \zeta} } 
  \newcommand{\ri}{ \frac{1}{R^2} }
  \newcommand{\fr}{ \lp \R \rp}
  \newcommand{\frt}{ \lp \R, \T \rp}
  \newcommand{\frtz}{ \lp \R,\T,\Z \rp}

  \newcommand{\fluxav}[1]{\la #1 \ra}
  \newcommand{\thetaav}[1]{\la #1 \ra_\T}

\newcommand{\cramplist}{
        \setlength{\itemsep}{0in}
        \setlength{\partopsep}{0in}
        \setlength{\topsep}{0in}}
\newcommand{\cramp}{\setlength{\parskip}{.5\parskip}}
\newcommand{\zapspace}{\topsep=0pt\partopsep=0pt\itemsep=0pt\parskip=0pt}

\section{Introduction}
\label{sec:intro}

The quiescent H-mode (QH-mode) tokamak regime addresses several burning-plasma
requirements of ITER/DEMO operation~\cite{garofalo15}.  To focus on one
favorable property in particular, QH-mode is an operation regime without  the
impulsive heat loads arising from edge-localized modes
(ELMs)~\cite{connor98,leonard06}.  The quiescent regime was first observed in
DIII-D~\cite{burrell05,solomon14,grierson15,garofalo15,burrell16,chen17}
discharges, but is also observed in JT-60U \cite{Sakamoto04,oyama05}, JET
\cite{solano10} and ASDEX-U \cite{suttrop05}.  As the mode activity associated
with QH-mode on DIII-D is characterized by small toroidal-mode numbers
($n_\phi\simeq1-5$), it is suitable for simulation with global MHD codes.  It
is hypothesized that transport associated with the low-$n_\phi$ perturbations
is enhanced during QH-mode leading to essentially steady-state profiles in the
pedestal region~\cite{Snyder07}.  The goal of the present work is to use the
extended-MHD model to develop further understanding of this hypothesis and
the experimental phenomenology.

Computational investigation of QH-mode provides many insights. Linear
computations with the M3D-C1 code show that low-$n_\phi$ modes are destabilized
by rotation and/or rotational shear while high-$n_\phi$ modes are
stabilized~\cite{chen16}. Nonlinear computations with the JOREK code
demonstrate the existence of a saturated state dominated by low-$n_\phi$
perturbations~\cite{liu15}.  In Ref.~\cite{King17}, initial analysis of
nonlinear NIMROD~\cite{Sovinec04} simulations of a QH-mode discharge with
broadband activity is presented. 

Our present work extends that work in three major ways: (1) As presented in
Sec.~\ref{sec:dynamics}, the saturated simulation, which uses steady-state
toroidal and poloidal flow as inferred from measurements, is compared to a
simulation with identical initial perturbations and profiles with the exception
that the steady-state flow is set to zero.  This simulation with flow
progresses to a low-$n_\phi$ saturated state whereas high-$n_\phi$ dynamics
dominate the simulation without flow.  (2) The flux-surface-averaged transport
is analyzed in Sec.~\ref{sec:transport}. It is established that fluctuation
amplitudes and phases lead to larger edge convective particle transport
relative to the thermal transport. This is similar to experimental observations
of density pump-out during QH-mode where the particle transport is larger than
the thermal transport. (3) Finally, the simulated rotation frequency as
computed by a synthetic diagnostic probe within the pedestal region is compared
to an analysis of the experimental signal from magnetic coils in
Sec.~\ref{sec:freq}.  The simulated perturbations rotate approximately at the
frequency of the steady-state ion flow, whereas slower rotation is observed in
the experiment which indicates the importance of effects beyond the
single-fluid MHD model. In particular, it is hypothesized that two-fluid
modeling is required to resolve this discrepancy.

\begin{figure}
  \includegraphics[width=8cm]{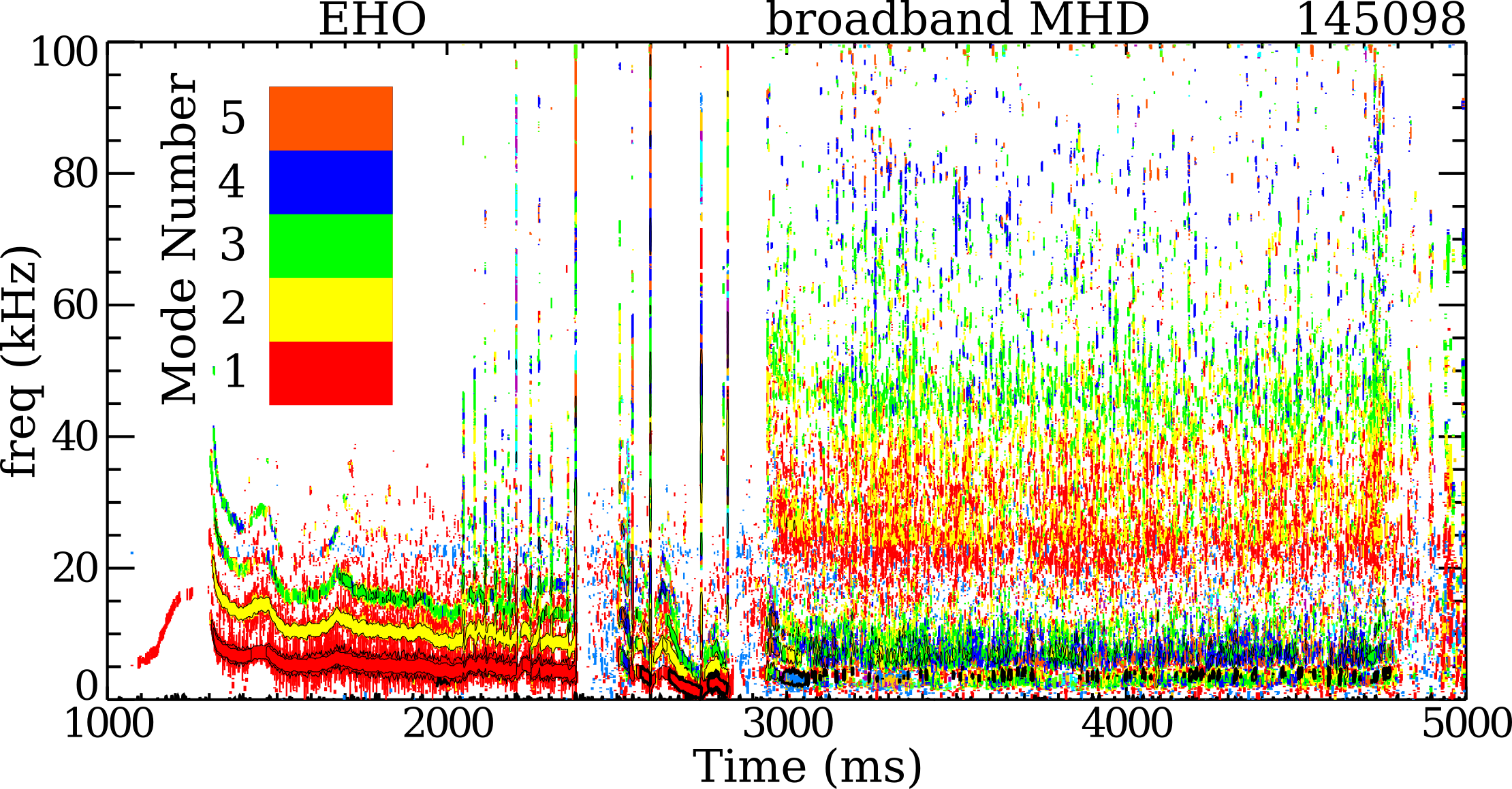}
  \caption{[color online]
  The cross-power spectrum plot from magnetic-probe
  measurements from DIII-D shot 145098 shows both an initial phase
  containing coherent EHO fluctuations, and a subsequent phase with
  a mix of broadband-MHD and coherent activity.  
  }
  \label{fig:newspec}
\end{figure}

NIMROD simulations are performed on a DIII-D QH-Mode discharge which exhibits
both a phase of coherent edge-harmonic oscillation (EHO) activity and a phase
of broadband MHD activity.  The cross-power spectrum plot from magnetic probe
measurements from DIII-D shot 145098 is shown in Fig.~\ref{fig:newspec}.  The
initial phase of the discharge is dominated by an EHO and later, as the input
neutral-beam torque is decreased, a subsequent phase with a mix of
broadband-MHD and coherent activity is observed. Relative to QH-mode operation
with EHO, operation with broadband MHD tends to occur at higher densities and
lower rotation and thus may be more relevant to potential ITER discharge
scenarios.  We analyze the shot at t=4250 ms where the broadband MHD state has
existed for over 1 second. This time period is chosen as it is a low-torque
part of the discharge that is relevant to ITER.  This is a lower-single-null
discharge that predates the discovery of the double-null wide-pedestal QH-mode
discharges~\cite{burrell16,chen17}.  Before proceeding to the results of our
analysis, we discuss the numerical formulation of the problem in the next
section (Sec.~\ref{sec:formulation}).
\section{Numerical formulation of the problem}
\label{sec:formulation}


The initial 2D condition is a Grad-Shafranov equilibrium, which is a
reconstruction of the plasma state as constrained by measurements. A problem
arises near the edge as QH-mode discharges typically have large current near
the separatrix.  \textsc{EFIT}\cite{lao85}, the most-widely used code for equilibrium
reconstructions, does not ordinarily include current in the scrape-off-layer
(SOL).  Without SOL current, reconstructions must have one of two undesirable
properties: 1) either there is an artificial constraint on the current where it
must smoothly vanish at the separatrix, or 2) there is a current discontinuity
at the separatrix.  The first constraint leads to incorrect gradients in the
pedestal region, and the second constraint leads to poor convergence for
nonlinear runs.  Most importantly, neither constraint gives SOL profiles that
are consistent with MHD-force balance.

To ameliorate these issues, our equilibrium solver,
\textsc{nimeq}~\cite{Howell14}, is extended to include SOL profiles and
current.  The profiles are extended from the Last Closed Flux Surface (LCFS)
into the SOL region. The SOL profiles for this particular case are described in
detail in Ref.~\cite{King17}. In Ref.~\cite{King16sol}, a quantitative analysis
of the impact of adding SOL current to the equilibrium is presented.
Re-solving the equilibrium and including SOL current and flow has little
effect on the linear analysis, but is critical for nonlinear simulations as it
eliminates discontinuities in current and flow at the LCFS.  These
discontinuities produce numerical problems when perturbations are advected over
the LCFS.  We present details of the configuration of these nonlinear
simulations next.


As seen in Figure~\ref{fig:newspec}, a time late in the discharge (4250 ms) is
chosen to analyze.  The simulation therefore starts from a state that is itself
a result of the low-$n_\phi$ perturbation activity associated with QH-mode.
The goal of our simulations is therefore {\em not} to understand the onset
phase of the broadband activity, but rather is to answer the question: ``Is the
observed low-$n_\phi$ activity calculated by extended MHD consistent with
experimental observations?''  Because there is some ambiguity in separating
external sources and the radial transport that is caused by both macroscopic
and microscopic turbulence, the simulations are run such that the sources
exactly maintain the equilibrium.  Although the simulation is run as an
initial-value simulation, it is the (possibly turbulent) steady-state that is
of interest.  Our simulation formulation is more similar to gyrokinetic
turbulence simulations and has similar concerns in terms of statistical
validation and temporal convergence issues~\cite{Holland:2011ef}.  One
difference with gyrokinetic simulations is that the modes are allowed to
generate $n=0$ transport.  As long as this transport does not induce changes to
the profiles that are far outside the measured constraints, our formulation is
accurate and self-consistent.

The simulation is initialized from a linear computation of modes with a
restricted toroidal-mode-number range ($n_\phi=1-8$). The mode energies
at $t=0 s$ are small, and the largest energy is contained within the
$n_\phi=4$ mode that has a spectral kinetic-energy content of
$4.2\times10^{-5}\;J$ and a spectral magnetic-energy content of
$4.4\times10^{-6}\;J$.  For the nonlinear runs, a resistive-MHD model
with an anisotropic stress tensor and anisotropic thermal heat conduction is used as the
computational resources required for two-fluid simulations are currently
prohibitive.  Although two-fluid nonlinearities could possibly change
the nonlinear cascade, the macroscopic nature of this mode means that the MHD model may be
sufficiently valid to begin gaining intuition in this regime. As
discussed later, two-fluid effects are likely critical for more
quantitative understanding. 

The resistivity profile is chosen such that the Lundquist number, $S$, in the
core is $1.1\times10^6$.  This choice of resistivity is enhanced by a factor of
100 relative to the Spitzer value for computational practicality.  The model
includes large parallel and small perpendicular diffusivities in the momentum
and energy equations.  Our simulations use $\chi_{\parallel}=10^8\;m^2/s$.  The
small perpendicular diffusivities are modeled as isotropic particle, momentum
and thermal diffusivities with a magnitude of $1\;m^2/s$.  Both the resistivity
and isotropic viscosity profiles are proportional to $T_e^{-3/2}$.  The
simulation is performed with a $60\times128$ high-order (bi-quartic)
finite-element mesh packed around the pedestal region to resolve the poloidal
plane and $24$ Fourier modes in the toroidal direction.  The boundary
conditions, on both the inner annulus and outer wall, are no-slip for the
velocity, Dirichlet for the density and temperature and a perfectly
conducting-wall boundary condition for the magnetic field.  Linear computations
show that the mode growth rates are unaffected by presence of the inner
boundary. For our nonlinear computations, the Dirichlet boundary conditions on
density and temperature provide an unconstrained source that prevents the edge
modes from simply transporting all the stored energy out of the confined plasma
region.
\section{Simulation dynamics with and without flow}
\label{sec:dynamics}

Experimentally, it is known that access to the QH-mode regime requires control
of the flow profile~\cite{garofalo11}. In particular, large
$\mathbf{E}\times\mathbf{B}$ flow shear is correlated with QH-mode operation.
In order to ascertain the impact of flow on the nonlinear evolution, we perform
simulations with and without steady-state flow.

\begin{figure}
  \includegraphics[width=8cm]{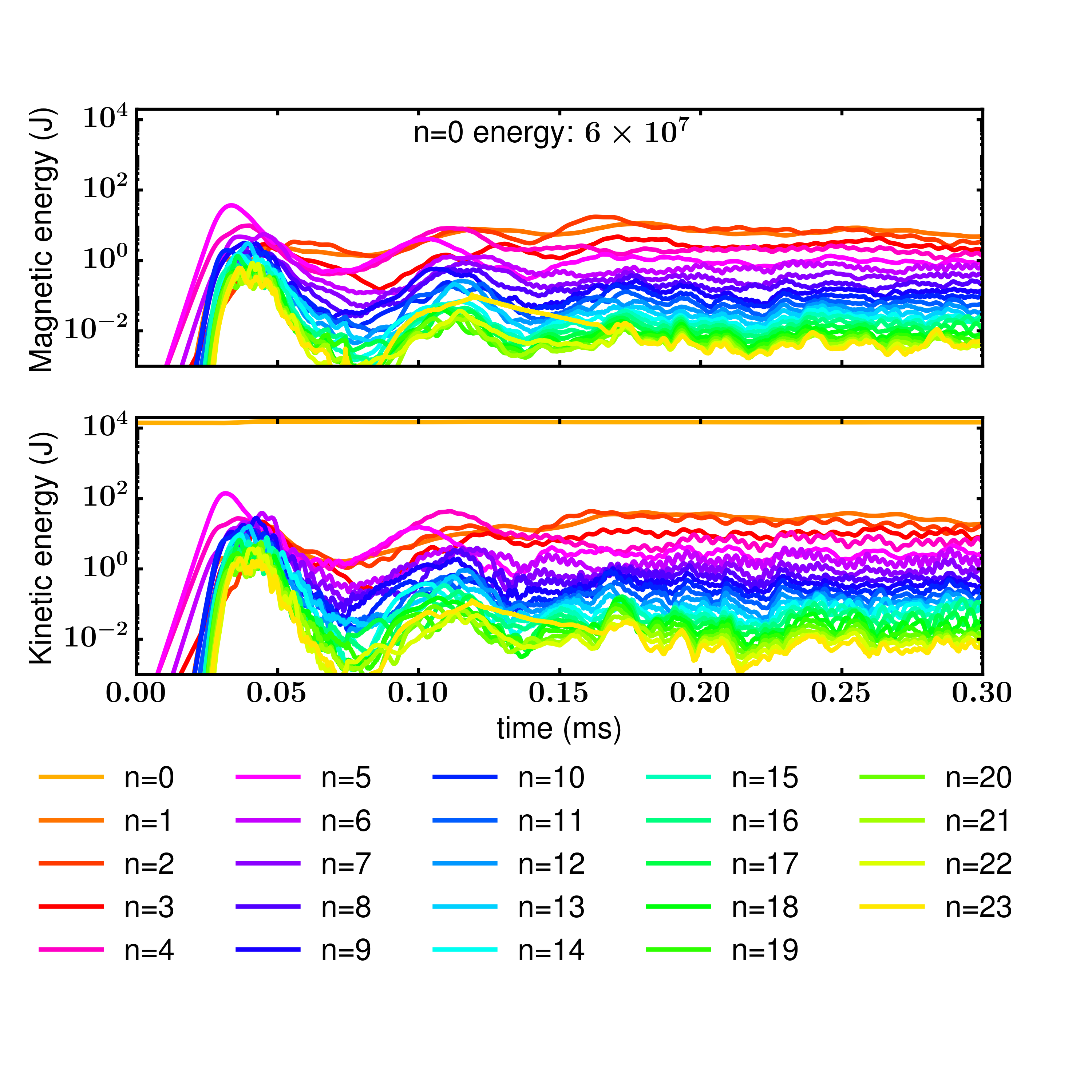}
  \caption{[color online]
    The time history of the magnetic (top) and kinetic (bottom)
    energies as decomposed by toroidal mode for a nonlinear simulation
    {\em with} steady-state flow. The simulation ultimately results
    in a low-$n_\phi$ saturated state.
  }
  \label{fig:energy}
\end{figure}

The energy evolution from the nonlinear simulation with steady-state flow, as
decomposed by toroidal mode number, of DIII-D QH-mode shot 145098 at $4250\;ms$
is shown in Fig.~\ref{fig:energy}.  This simulation includes the toroidal and
poloidal rotation profiles inferred from the Carbon impurity species as measured by
Charge-Exchange Recombination (CER) spectroscopy.  The simulations are
initially dominated by a $n_\phi=5$ perturbation that saturates at around
$30\;\mu s$.  After this time a nonlinear saturated quasi-turbulent state
develops and the $n_\phi=1$ and $2$ perturbations become dominant through an
inverse cascade.  From $0.15ms$ onwards, modulations are observed in the energy
(particularly the kinetic energy) and there is continued interplay between the
perturbations. We use the phrase quasi-turbulent as the state is not laminar
but it also is not strongly turbulent. If this were a laminar state such as a
typical tearing mode case, the perturbed energies would be well-separated and
constant in time.   

\begin{figure}
  \includegraphics[width=8cm]{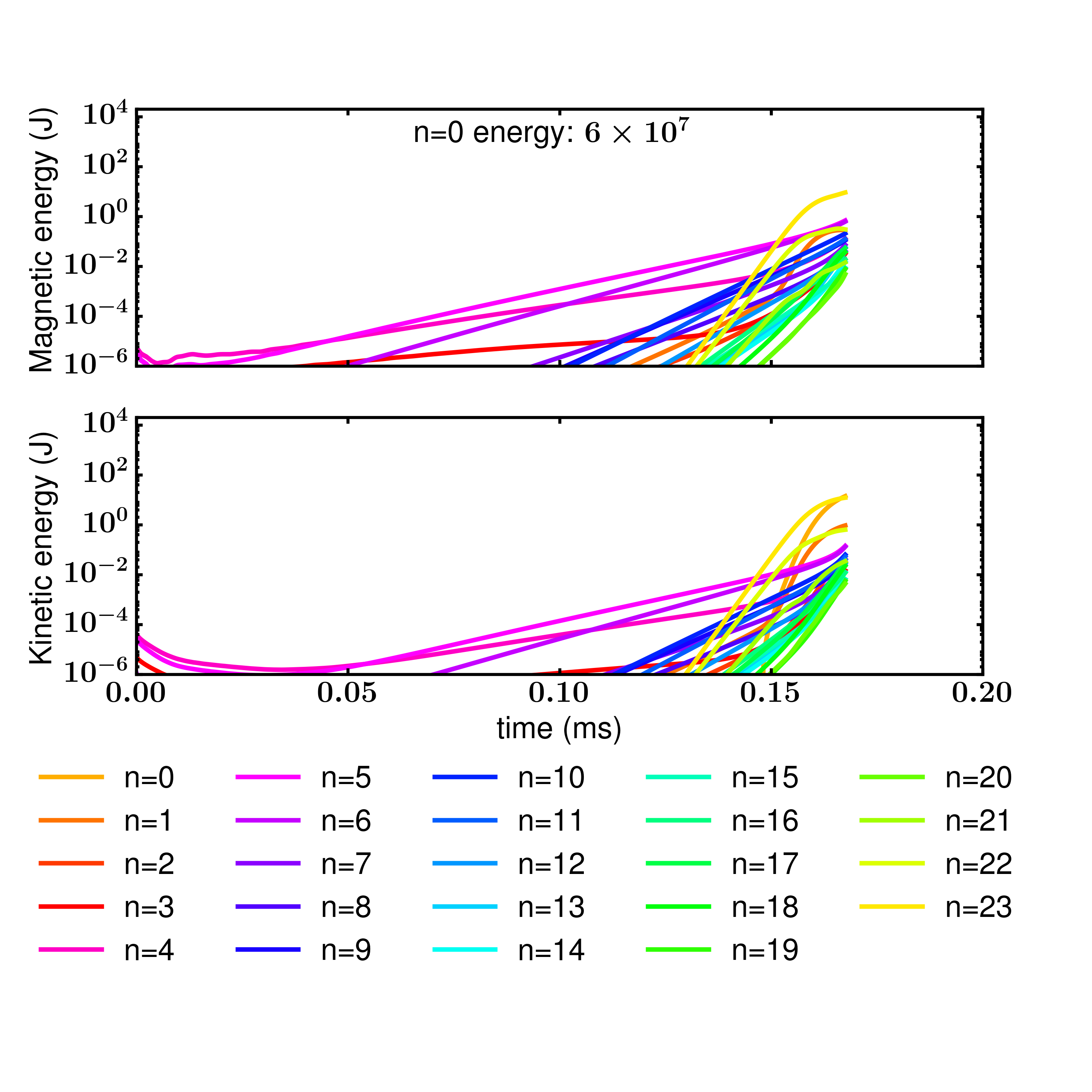}
  \caption{[color online]
    The time history of the magnetic (top) and kinetic (bottom) energies as
decomposed by toroidal mode for a nonlinear simulation {\em without}
steady-state flow.  The dynamics ultimately proceed to high-$n_\phi$ and the
computation stops when the limit of the spatial resolution is reached.
  }
  \label{fig:energy-noflow}
\end{figure}

Simulations {\em without} steady-state flow observe an ELM-like high-$n_\phi$
evolution.  With an identical initial condition as the simulations with
flow, the low-$n_\phi$ modes are initially stable and later are
slowly growing. The dynamics are ultimately overtaken by fast growing
high-$n_\phi$ modes as shown in Fig.~\ref{fig:energy-noflow}. The computation
stops when the limit of the spatial resolution is reached.  This
behavior is similar to extended-MHD simulation of ELM
dynamics~\cite{pankin07,sugiyama10,huijsmans13}, not
QH-mode. Importantly, the result that simulations with steady-state flow
(and the associated large flow shear) saturate to a low-$n_\phi$ state
while simulations without steady-state flow (and thus no flow shear)
lead to dynamics at high-$n_\phi$ is consistent with the experimental
observation that QH-mode access requires large-edge flow shear.

\begin{figure*}
  \includegraphics[width=16cm]{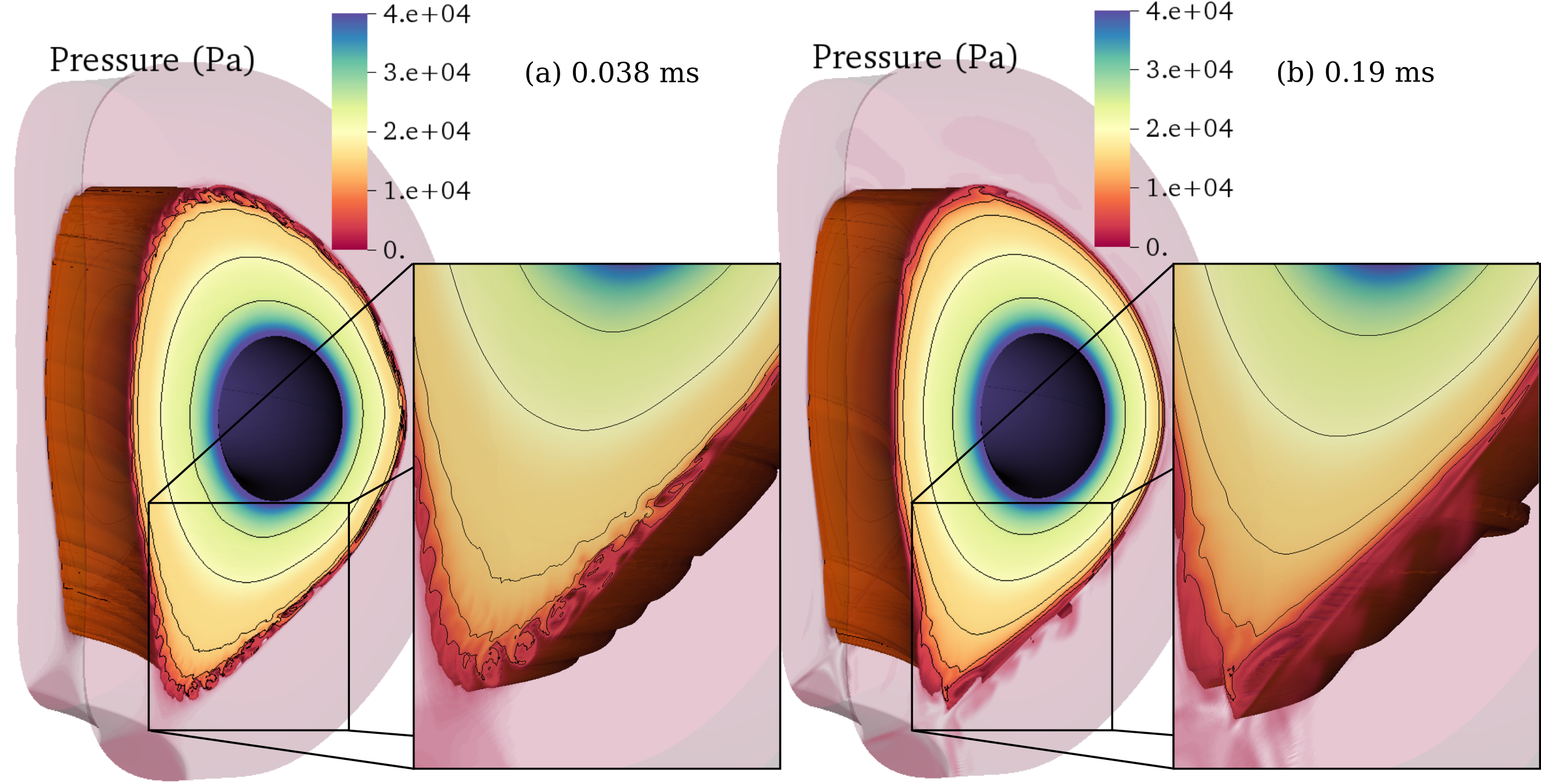}
  \caption{[color online]
   3D pressure contours as a cut of the full 3D torus at two different times
demonstrate the nonlinear evolution.  (a) During the initial stages (t=$38\;\mu
s$), dominantly $n_\phi=5$ eddies of hot, high density plasma are ejected from
the pedestal and are advected poloidally in the counter-clockwise direction
while being sheared apart.  (b) Later in time (t=$0.19\;ms$), these eddies are
sheared apart and a quasi-turbulent steady-state develops that is dominated by
low-$n_\phi$ perturbations. Fingers of hot plasma form near the x-point in the
magnetic-flux expansion region; these are advected poloidally in the
counter-clockwise direction and are ultimately sheared apart as they reach the
outboard midplane. 
\href{https://www.youtube.com/watch?v=8m-pASKQcMw}{(Multimedia view)}
}
  \label{fig:pres3D}
\end{figure*}

We turn to analysis of the simulation {\em with} steady-state flow for the
remainder of the paper.  The eddies from the simulation with flow can clearly
be seen in Fig.~\ref{fig:pres3D}. This figure plots 3D pressure contours on
a poloidal cut of the full 3D torus at two different time slices to demonstrate
the nonlinear evolution.  During the initial stages (t=$38\;\mu s$), dominantly
$n_\phi=5$ eddies of hot, high-density plasma are ejected from the pedestal and
are advected poloidally in the counter-clockwise direction while being sheared
apart through interaction with both the sub-dominant $n_\phi=4$ perturbation
and the underlying sheared flow.  After this time the perturbation amplitudes
are reduced and plots show a smoke-like off-gassing behavior (not shown in
figure). Later in time (t=$0.19\;ms$), a quasi-turbulent steady-state develops
that is dominated by low-$n_\phi$ perturbations. Fingers of hot plasma form near the
x-point in the magnetic-flux expansion region, these are advected poloidally in
the counter-clockwise direction and are ultimately sheared apart as they reach
the outboard midplane similar to theoretical considerations ~\cite{Guo15}. 
Figure~\ref{fig:pres3D}(b) shows both a recently formed
finger near the x-point and a highly sheared eddy near the outboard midplane.
A video of these dynamics is available with this article.  The period of
modulation in the spectral energies in Fig.~\ref{fig:energy} is consistent with
the time it takes to form and shear apart the eddies shown in
Fig.~\ref{fig:pres3D}(b).
\section{Transport analysis}
\label{sec:transport}

\begin{figure}
  \includegraphics[width=8cm]{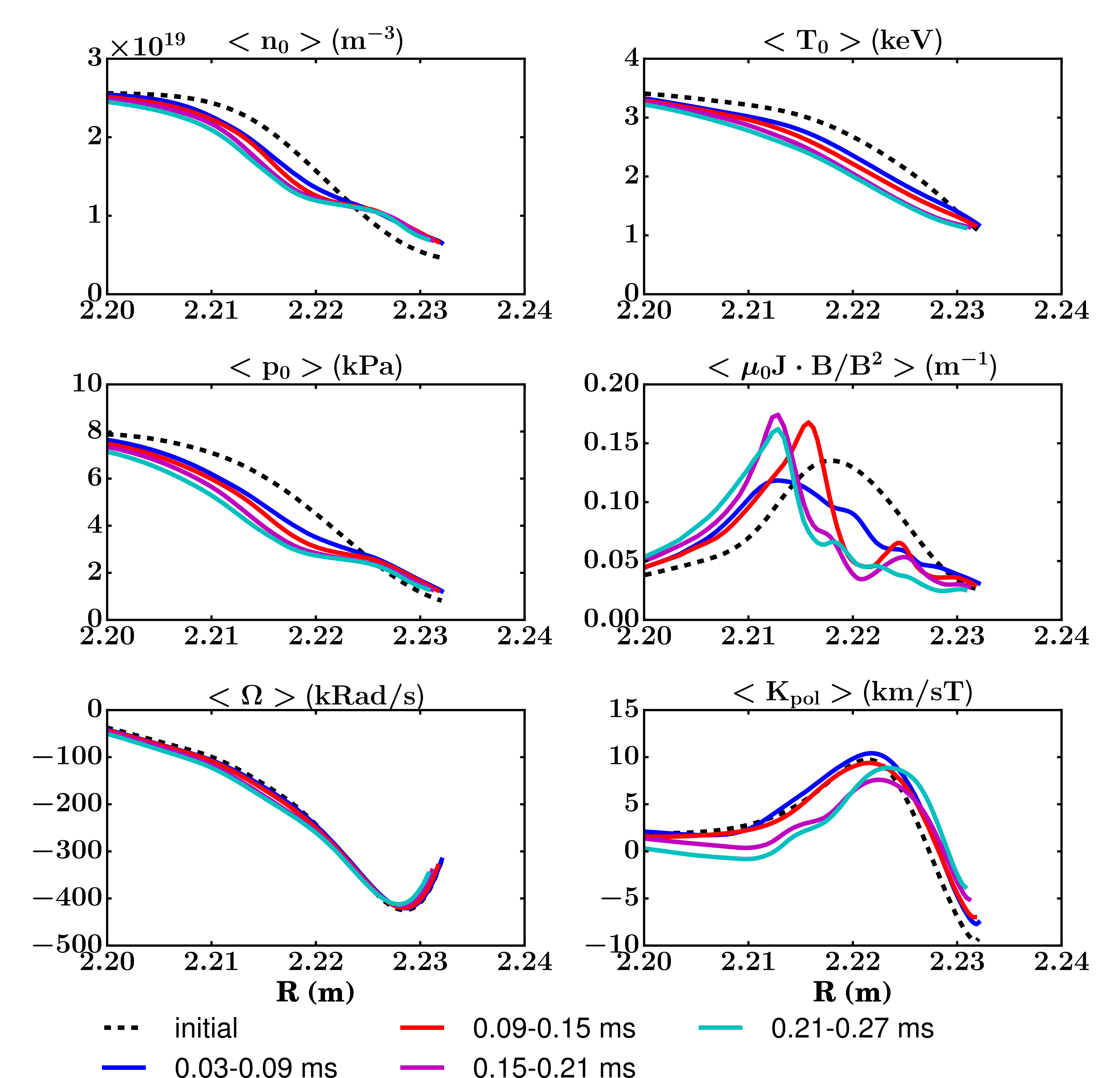}
  \caption{[color online]
  Flux-surface-averaged values of the density, temperature,
  pressure, parallel-current density, toroidal rotation frequency and
  poloidal rotation neoclassical coefficient relative to the location
  in the pedestal region on the outboard midplane. 
  The initial condition and average values during four time windows
  are shown.
  }
  \label{fig:profs}
\end{figure}

Having shown that low-$n_\phi$ perturbations lead to near steady-state
perturbations within the pedestal, it is natural to study the saturation
mechanisms and transport in more detail.  Figure~\ref{fig:profs} shows the
flux-surface-averaged values of the density, temperature, pressure,
parallel-current density, toroidal rotation frequency and poloidal-rotation
neoclassical coefficient relative to the location in the pedestal region on the
outboard midplane for the initial condition and average values during four time
windows. The symmetric flux surfaces associated with the 2D magnetic field are
used to determine the surfaces in the flux-surface averages; i.e., we include
both the equilibrium and the perturbed $n_\phi=0$ contributions. The profiles during
the latter two time windows, as associated with the fully saturated state
($>0.15\;ms$ in Fig.~\ref{fig:energy}), are very similar indicating a
time-averaged steady 3D solution. The modifications to the flow profiles are
small relative to the transport of pressure and current.  We thus conclude that
the saturation mechanism is related to a flattening of these latter profiles,
and not changes to the flow profiles. This turbulent state leads to larger
particle transport relative to thermal transport as shown by more flattening of
the density profile relative to the temperature profile in
Fig.~\ref{fig:profs}, qualitatively consistent with the experimental
observations of density pump-out during QH-mode~\cite{garofalo15}. This result
is surprising as the magnetic field is stochastic within the pedestal (see
Ref.~\cite{King17}). With the highly anisotropic thermal conduction as included
in the model, one would naively expect that thermal transport should be
enhanced relative to the particle transport instead of the reverse.  

\begin{figure}
  \includegraphics[width=6cm]{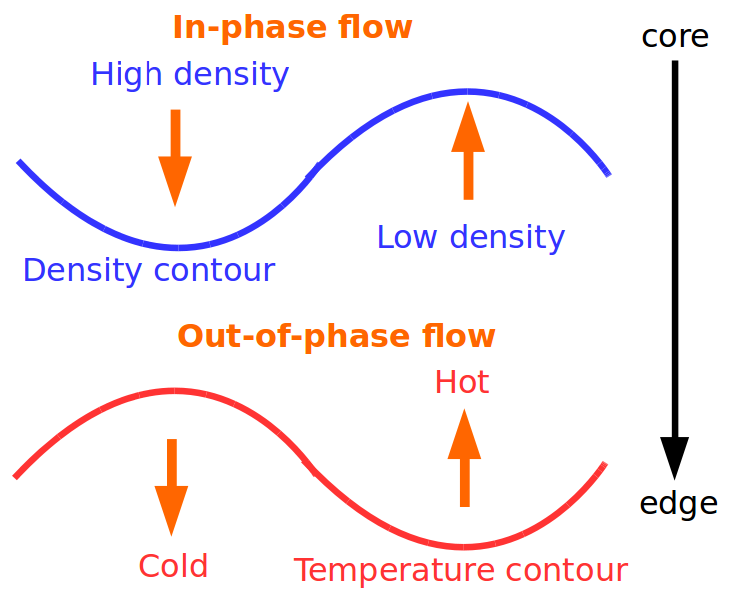}
  \caption{[color online]
    A schematic showing in-phase flow (arrows) with a density contour and
out-of-phase flow (arrows) with a temperature contour.
  }
  \label{fig:phase}
\end{figure}

The transport for density is advective, while temperature includes both
convection and anisotropic thermal conduction.  The contravariant
normal component of the convective flux through a surface, $\Gamma_f$, is the
flux-surface integral of the product of the density and temperature
fluctuations, $\tilde{f}$, and the velocity normal to the surface,
\begin{equation}
\Gamma_f = \int \frac{\tilde{f} \tilde{\mathbf{v}}}{\mathbf{B}_0\cdot\nabla\theta} \cdot \mathbf{dS} 
/ \int \frac{dS}{\mathbf{B}_0\cdot\nabla\theta}\;.
\end{equation}
Here $\mathbf{B}_0$ is the $n_\phi=0$ magnetic field and $\theta$ is a poloidal
coordinate.  For this quadratic quantity, only mode self-interaction
contributes after the application of a toroidal average to the toroidally
symmetric surfaces.  Importantly for $\Gamma_T$ and $\Gamma_n$, only the phase
and the amplitude of the perturbation relative to the normal velocity may
impact the overall normalized magnitude. In Fig.~\ref{fig:phase}, a schematic
shows how an in-phase density perturbation leads to enhanced outward (core
to edge) transport while an out-of-phase temperature perturbation leads to
inward (edge to core) transport. For mixed phases, the overall transport may
vanish.

\begin{figure*}
  \includegraphics[width=16cm]{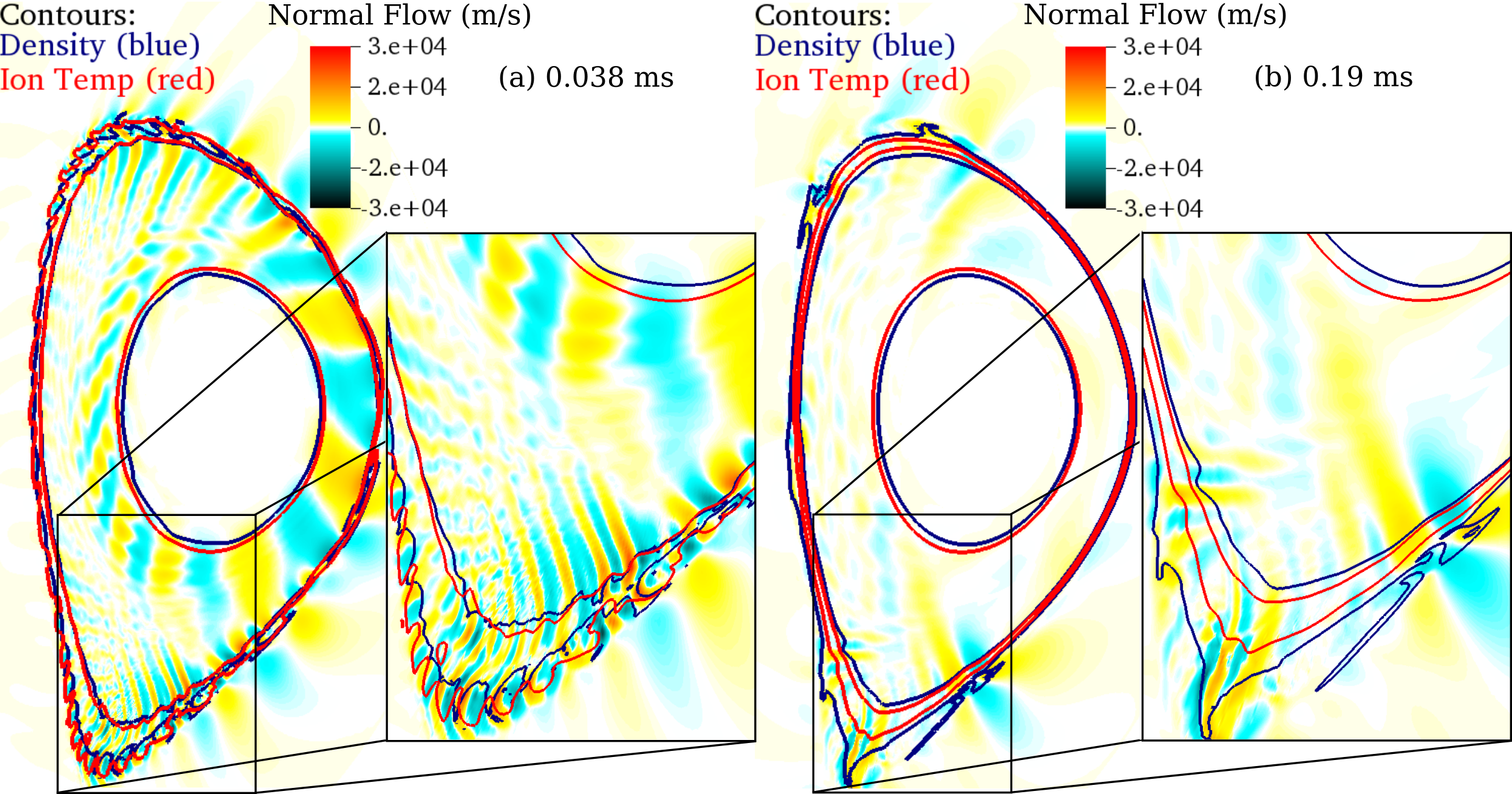}
  \caption{[color online]
  A poloidal cut of the tokamak at (a) $38\;\mu s$ and (b) $0.19\;ms$ showing
three contour lines each of density (blue) and temperature (red) overlaying
pseudocolor contours of flow normal to the $n_\phi=0$ magnetic field (note the
normal flow color scale is adjusted in such a way as to show the global
structure of the localized perturbations).  Early in time (a), density is in
phase with the normal flow while the temperature is out of phase. This effect
leads to larger density flux relative to the thermal flux.  Later in time (b),
a more mixed result is achieved as both the density and temperature are more in
phase with the flows, but the large variation in the density contours shows
that the amplitude of the density perturbation is larger than that of the
temperature perturbation. 
\href{https://www.youtube.com/watch?v=XeKNAJ14jrs}{(Multimedia view)}
  }
  \label{fig:ntvn2D}
\end{figure*}

Figure~\ref{fig:ntvn2D} shows temperature and density contours super-imposed
with flux-surface-normal velocity color-plot on a poloidal cut of the tokamak
at early and late times during the simulation.  Early in time
(Fig.~\ref{fig:ntvn2D}(a)), density is in phase with the normal flow while the
temperature is out of phase. This effect leads to larger density flux relative
to the thermal flux.  Later in time (Fig.~\ref{fig:ntvn2D}(b)), a more mixed
result is achieved as both the density and temperature are more in phase with
the flows, but the large variation in the density contours shows that the
amplitude of the density perturbations is larger than that of the temperature
perturbations.

\begin{figure}
  \includegraphics[width=8cm]{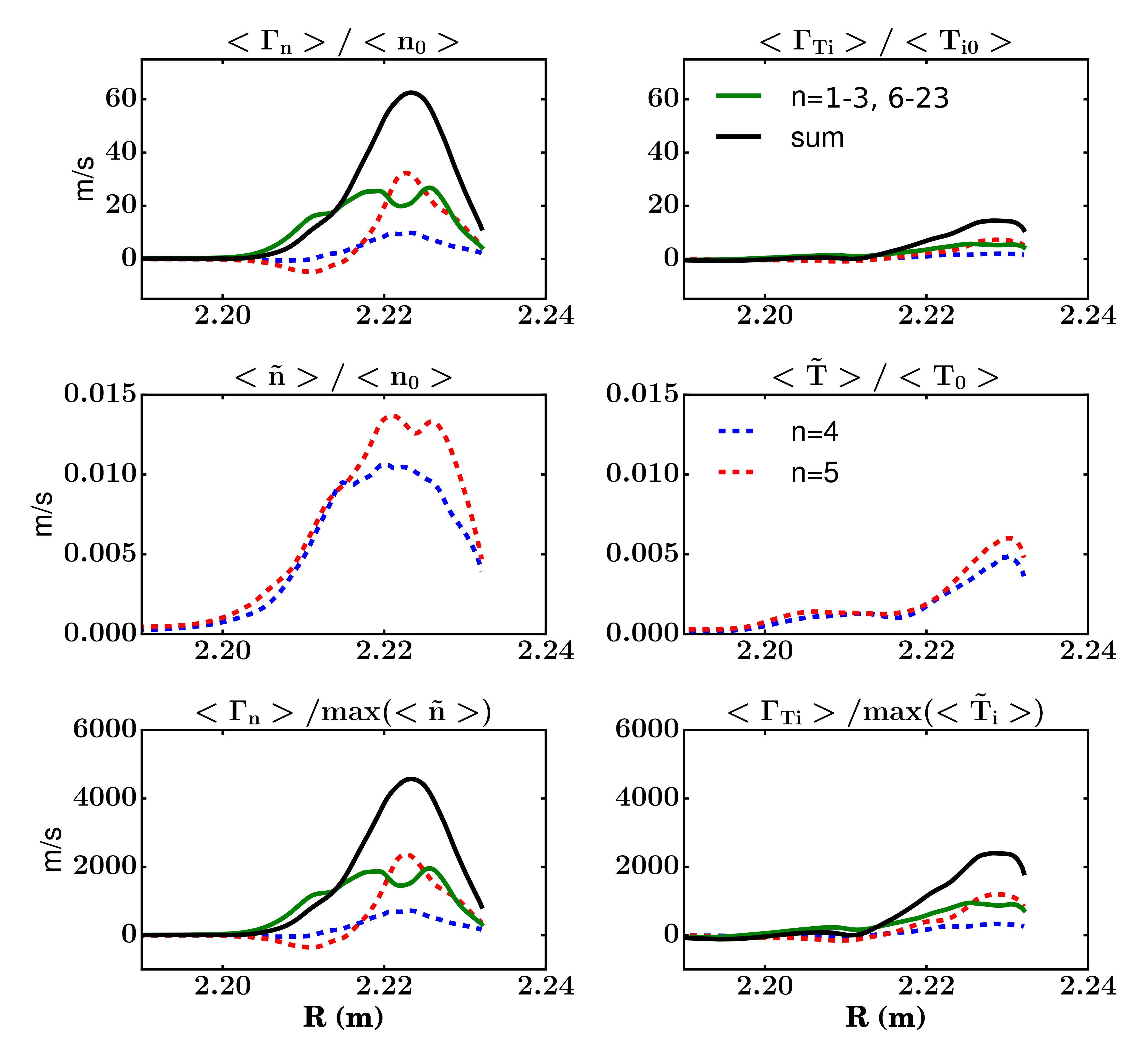}
  \caption{[color online]
  Flux-surface- and time-averaged ($0.03<t<0.09\;ms$) transport ($\Gamma$),
  total and broken out by contributions, for
  density and temperature as normalized to the $n_\phi=0$ value at $R=2.22\;m$
  on the outboard midplane (top row) and to the maximum perturbation amplitude
  (bottom row). The middle row displays the flux-surface- and time-averaged
  density and temperature perturbations. Only the modes associated with the 
  largest perturbations are displayed.
  }
  \label{fig:transA}
\end{figure}

\begin{figure}
  \includegraphics[width=8cm]{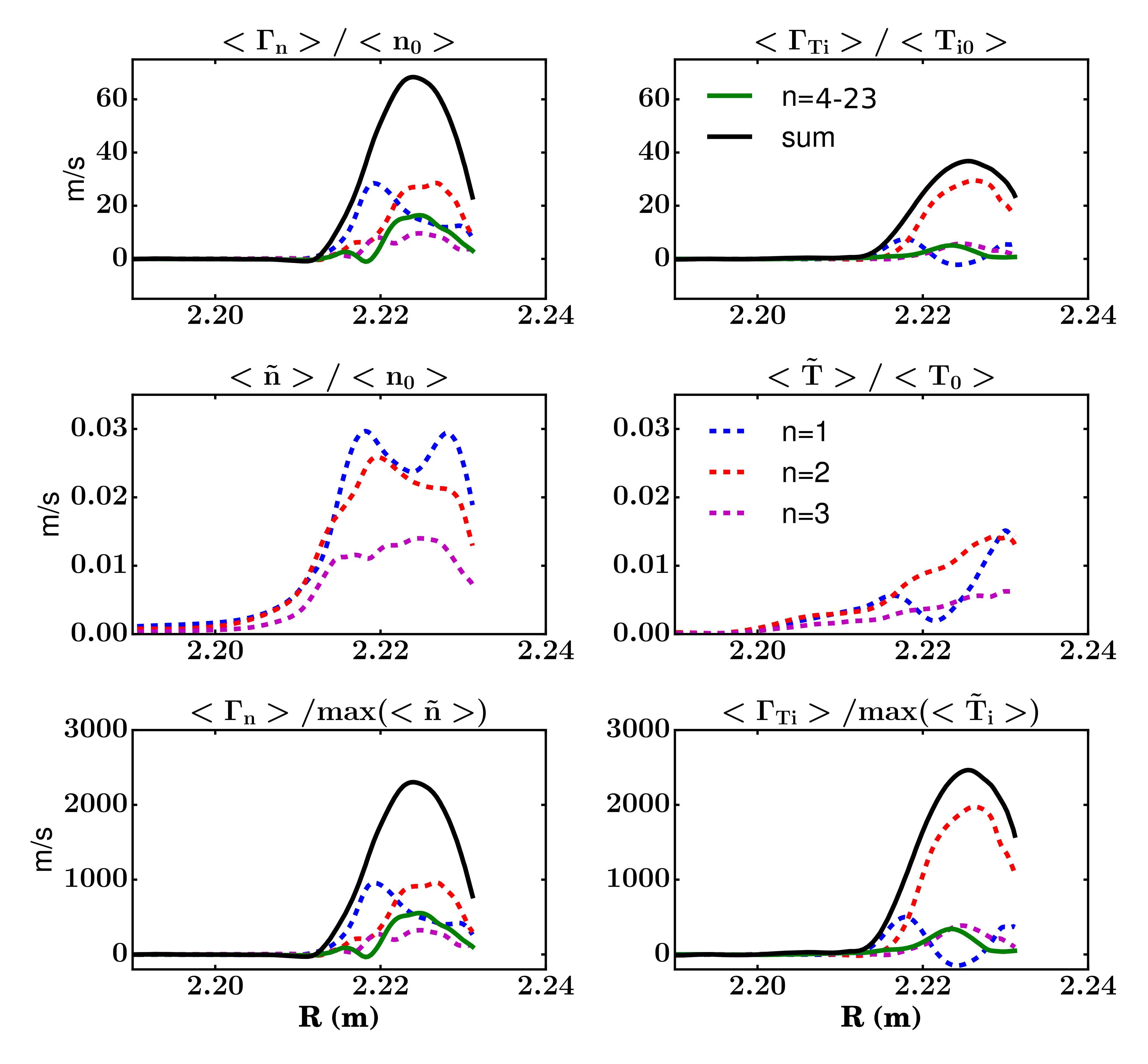}
  \caption{[color online]
  Flux-surface- and time-averaged ($0.16<t<0.21\;ms$) transport ($\Gamma$),
  total and broken out by contributions, for
  density and temperature as normalized to the $n_\phi=0$ value at $R=2.22\;m$
  on the outboard midplane (top row) and to the maximum perturbation amplitude
  (bottom row). The middle row displays the flux-surface- and time-averaged
  density and temperature perturbations. Only the modes associated with the 
  largest perturbations are displayed.
  }
  \label{fig:transC}
\end{figure}

A more rigorous analysis is performed with Figs.~\ref{fig:transA} and
\ref{fig:transC}. This analysis plots the flux-surface-integrated normalized
fluxes and flux-surface-averaged perturbation amplitudes for both density and
temperature in time-averaged early (Fig.~\ref{fig:transA}, $0.03<t<0.09\;ms$)
and late (Fig.~\ref{fig:transC}, $0.16<t<0.21\;ms$) time windows. Both the
total flux and contributions to it from the dominant modes during each time
window are shown. The top two figures of each plot show the overall density
and temperature convective transport during each window as normalized by the
$n_\phi=0$ values at $R=2.22\;m$ on the outboard midplane. In both time
windows, the density transport is large relative to the thermal transport. It
is larger by factors of approximately $6\times$ and $2\times$ in the early and
late windows, respectively.  The second row of plots compares the normalized
perturbation amplitudes, again normalized to the $n_\phi=0$ values at
$R=2.22\;m$.  Consistent with the toroidally decomposed global energy,
Fig.~\ref{fig:energy}, the $n_\phi=4$ and $5$ perturbations are dominant in the
early time window and the $n_\phi=1-3$ perturbations are largest during the
late time window.  For both time windows, the density perturbations are larger
than the temperature perturbations by factors of $3\times$ to $2\times$ for
the early and late windows, respectively. Thus the difference in amplitude
fully accounts for the difference in the fluctuation-induced transport in the
late time window, while the early time window still has a factor of
approximately $2\times$ that is unaccounted for. The last row of plots shows
the convective transport normalized by the maximum perturbation amplitude. This
choice of normalization provides an amplitude-independent plot of the effect of
the phase. Consistent with estimates from earlier plots, the phase differences
contribute a factor of $2\times$ in the early time window while they are
negligible in the late time window. This more rigorous analysis shows that both
amplitude and phase differences contribute to large fluctuation-induced
convective density transport relative to the thermal transport.

More qualitatively, the differences in phase and amplitude arise from
differences in the underlying density and temperature equations.  The density
equation captures the effects of advection and compression and also includes a
small numerical diffusivity with coefficient $D_n$,
\begin{equation}
\frac{d n}{d t} = - n \nabla \cdot \mathbf{v}  + D_n \nabla^2 n \;.
\end{equation}
In contrast, the temperature equation differs in factors of the ratio
of specific heat, $\Gamma$, and through the inclusion of large anistropic thermal 
conduction,
\begin{equation} 
\frac{n}{\Gamma -1 } \frac{d T}{d t} = - n T \nabla \cdot \mathbf{v} + \nabla \cdot
n_0(\psi_0) \left( \chi_\perp + \chi_\parallel \hat{\mathbf{b}} \hat{\mathbf{b}} \cdot \right)
\nabla T \;.
\label{eq:T}
\end{equation}
The factor of $n(\psi_0)$ is the density on axis and is used to simplify 
the specification of the thermal conduction in units of $m^2/s$.
It can be inferred that differences in the temperature-evolution equation
result from differences with the factors of $\Gamma$, and, likely more
importantly, the inclusion of anisotropic thermal conduction. The additional terms modify the
amplitude and phase relative to the density and reduce the time-averaged
fluctuation-induced convective thermal transport when compared to the particle
transport.

Of course, the thermal transport is enhanced by anisotropic thermal conduction
along the stochastic-magnetic-field lines. As the analysis of the convective
transport in the preceding paragraphs appears to explain the differences in the
computed density and thermal transport, we expect the conductive transport to
be small. In order to ascertain the approximate level of transport we next
estimate the confinement times associated with the conductive and convective
transport.  Using the values of the convective transport during the late time
stage (Fig.~\ref{fig:transC}), the convective confinement times for density and
temperature are approximately 
\begin{equation}
\tau_{conv,n} \simeq \frac{ L_\perp <n_0>}{<\tilde{n}\tilde{v}_n>} \simeq 5\times 10^{-4}\;s
\end{equation}
and
\begin{equation}
\tau_{conv,T} \simeq \frac{ L_\perp <T_0>}{<\tilde{T}\tilde{v}_n>} \simeq 10^{-3} \;s,
\end{equation}
respectively. We estimate that $L_\perp=0.03\;m$ and $<n_0>$ and $<T_0>$
are evaluated at $R=2.22\;m$. Phase and amplitude effects are included in the
flux-surface-integrated expression in the denominator.  The parallel conductive
confinement time can be estimated based on the parallel length scale,
$L_\parallel$, from the magnetic-field puncture plot of Fig.~8 of
Ref.~\cite{King17} of $2500\;m$ as 
\begin{equation}
\tau_{cond,\parallel} \simeq \frac{L_\parallel^2 <n_0>}{\chi_\parallel n_0(\psi_0)} \simeq 1.5 \times 10^{-2}\;s,
\end{equation} 
which is longer than the convective confinement time. The factors of density
are added to produce an evaluation of the parallel thermal conduction in the
pedestal that is consistent with our temperature-evolution equation
(Eqn.~\eqref{eq:T}).  Our parallel conduction model is somewhat crude; however,
if we assume that the thermal conduction is in the fastest collisionless limit
of free-streaming along field lines, we find that the confinement time is then
\begin{equation}
\tau_{cond,\parallel,vTe} \simeq \frac{L_\parallel} {v_{Te}} \simeq 2\times10^{-3}\;s.
\end{equation}
Here a thermal velocity of $v_{Te}=\sqrt{k_B T_e/m_e}\simeq10^6\;m/s$ is used which assumes an
electron temperature of $10eV$.  This confinement time is comparable to the
convective thermal confinement time; however, it is unlikely that collisionless
conditions exist near the LCFS and thus the conductive confinement time is
likely greater.
\section{Frequency analysis}
\label{sec:freq}

\begin{figure}
  \includegraphics[width=8cm]{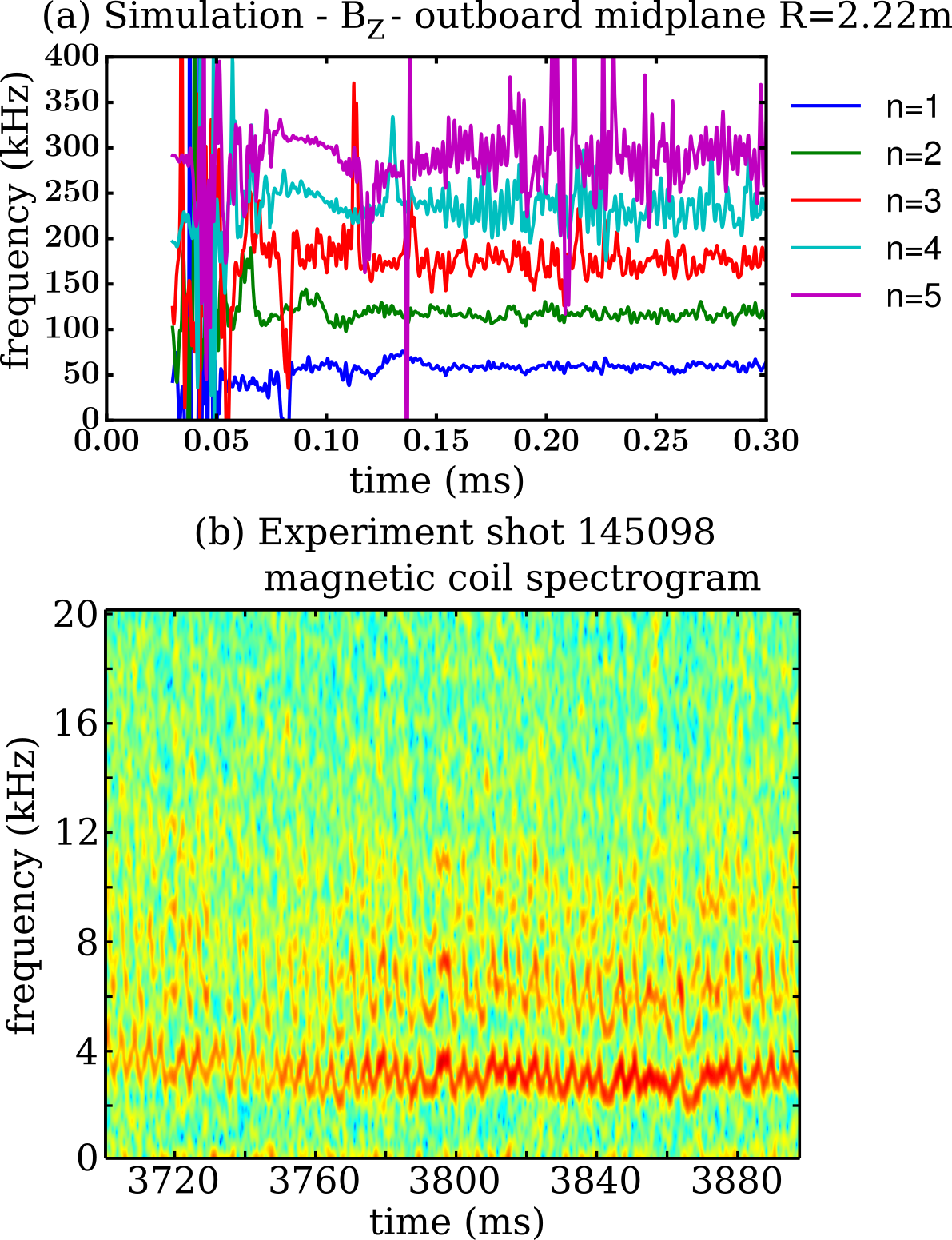}
  \caption{[color online]
(a) The frequency of the $\tilde{B}_Z$ perturbation as measured by a toroidal
set of synthetic probes at $R=2.22\;m$ on the outboard midplane and decomposed
by toroidal-mode number relative to (b) a frequency spectrogram produced from a
set of magnetic pickup coils at the wall in the experiment. Although we are
unable to assign mode numbers to the rotating structures in the experiment, it
is clear that the simulated perturbations rotate much faster than those in the
experiment.
  }
  \label{fig:freq}
\end{figure}

Figure~\ref{fig:freq} shows the frequency of the $\tilde{B}_Z$ perturbation as
measured by a toroidal set of synthetic probes at $R=2.22\;m$ on the outboard
midplane and decomposed by toroidal-mode number relative to a frequency
spectrogram produced from a set of magnetic pickup coils at the wall in the
experiment. Although we are unable to assign mode numbers to the rotating
structures in the experiment, which precludes comparison of the amplitude and
phase of the perturbations, it is clear that the simulated perturbation rotates
much faster than those in the experiment.  Our single-fluid computations rotate
at approximately the frequency associated with the toroidal ion flow.  It is
likely that two-fluid effects (e.g.  Ref.~\cite{Coppi64,King14,King16meudas}),
which are known to modify the frequency through the mediation of the
perturbations by both the ion and electron fluids are required to produce
dynamics with a more representative frequency. The ion and electron fluids
rotate in opposing directions for this discharge.  Another possibility is that
a drag on resistive wall components modifies the rotation frequency. This
effect is not captured by the perfectly conducting wall boundary condition
included in the simulation.
\section{Discussion}
\label{sec:discussion}

\subsection{Power flow}

As QH-mode leads to steady-state pedestal profiles, the fluctuation-induced
power flow through the pedestal can be bounded from the experimental data.
With the configuration of our present modeling, there is no prior constraint on
how much power the perturbations transport through the pedestal.  However, we
are unable to make this comparison given two limitations of the current model:
(1) it uses enhanced dissipation parameters and (2) it does not include Ohmic
and viscous heating. The former choice is made for computational practicality
while the latter is a natural consequence of the former as the heating rate
with enhanced dissipation would be too large. Computations with more realistic
dissipation parameters and heating require higher resolution and large
computational resources. Additionally, high-resolution computations are
currently limited by a time-step restriction that results from the inclusion of
flow in the steady state. This restriction requires the time-step size to
become smaller as the poloidal resolution is increased.  As this time-step
restriction is not inherent to NIMROD's discretization or implicit time-step
advance, further work is required to characterize its origin and circumvent it.

\subsection{Sheared eddy hypothesis for QH-mode perturbations}

Our simulations are consistent with qualitative features of DIII-D QH-mode;
however, further study requires validation with local
diagnostics~\cite{terry2008validation} to determine whether our simulated state
more closely resembles broadband MHD or coherent EHO. As explained in
Sec.~\ref{sec:freq}, it is expected that extensions to the modeling are
critical for quantitative comparison which likely requires the correct
frequency dependence.  

Despite this, the sheared nature of the computed saturated state leads to a
hypothesis on the nature of different QH-mode perturbations.  We posit that the
broadband state is driven similar to the coherent EHO; however, the
perturbations are sheared apart before they can reach a sufficient amplitude to
form a coherent structure. Additionally, we expect that for moderate flow shear
the eddy sizes are of the order of the ion gyroradius or larger, producing
rotation in the ion direction. As the shearing rate increases relative to the
growth of the instability, it restricts the saturated width of the
perturbations to a radial extent that is smaller or on the order of the ion
gyroradius. In this regime, the perturbation becomes dominated by the electron
dynamics, and it then rotates in the direction of the electron
fluid. This is consistent with experimental observations of the coherent EHO
rotating in the ion direction, and broadband QH-mode rotating in the electron
direction. Two-fluid computations that examine multiple cases with varied
low-$n_\phi$ perturbations are required to explore this hypothesis further.

\subsection{Limitations of modeling assumptions}

Ultimately, it is desired to have a modeling capability that can predict the
onset of QH-mode. In such a simulation, the evolution of the 2D fields that
comprise our underlying equilibrium are modeled. Presumably, a sufficient
demonstration evolves the 2D fields through the stability boundary associated
with the 3D modes and tracks the 3D dynamics through saturation. Our present
simulation, which assumes the underlying state is static in time, is not able
to capture this evolution. This requires a simulation that couples a model of
the underlying drives and transport (which is a challenging problem itself, see
e.g. Ref.~\cite{callen10}) that are outside of the present model coupled to the
dynamics and fluctuation-induced transport of the MHD model. This is left to
future studies.

\section{Summary}
\label{sec:summary}

The simulations shown here have demonstrated the capability of the extended-MHD
code NIMROD to model the low-$n_\phi$ perturbations associated with QH-mode.
With this capability, our longer term goal is to produce a model that is able
to determine the feasibility of QH-mode operation in future burning-plasma
devices. This goal requires more quantitative validation on a broad set of
discharges with varying parameters in addition to more focus on the role of the
flow shear in the accessibility criterion for QH-mode.

Our simulations that include the flow profiles inferred from CER Carbon
impurity measurements produce a quasi-turbulent state dominated by the
low-$n_\phi$ perturbations. This state is only achieved after careful inclusion
of the reconstructed equilibria including the addition of SOL profiles and
associated current. The interplay between driven low-$n_\phi$ perturbations and
the flow shear prevents the formation of laminar dynamics. Importantly,
simulations without flow have dynamics that develop at high-$n_\phi$ and do not
saturate consistent with experimental observations on the importance of large
edge-flow shear to QH-mode access.

Analysis of the transport generated by the low-$n_\phi$ perturbations shows
that a flattening of the pressure and current profiles is the
likely mechanism that eliminates the free-energy drive and leads to saturation.
Regarding the modifications to the pressure profile, the particle transport is
large relative to thermal transport. Further analysis shows that differences in
the particle and thermal fluxes result from a difference in the amplitudes of
the density and temperature perturbations as well as phase decorrelation of the
temperature with respect to the normal flow.  The conclusion is that
differences in the temperature-evolution equation resulting from the inclusion
of the anisotropic thermal conduction modify the amplitude and phase relative
to the density perturbation and reduce the time-averaged convective thermal
transport when compared to the particle transport.

\vspace{-0.200in}
\section*{Acknowledgements}
\vspace{-0.100in}

The authors would like to thank the NIMROD team, the Center for Extended
Magnetohydrodynamic Modeling and the DIII-D
collaboration. In particular we would like to thank Richard Buttery, Jim Callen,
Chris Hegna, Eric Held, Tobin Munsat, Chris Muscatello, Carl Sovinec, Linda
Sugiyama, and Theresa Wilks for excellent discussions and comments. We would
also like to thank Boris Breizman for stimulating our interest in computing the
power flow.  This material is based on work supported by US Department of
Energy, Office of Science, Office of Fusion Energy Sciences under award numbers
DE-FC02-06ER54875$^1$, DE-FC02-08ER54972$^1$ (Tech-X collaborators), and
DE-FC02-04ER54698$^2$ (General Atomics collaborators).  This research used
resources of the Argonne Leadership Computing Facility, which is a DOE Office
of Science User Facility supported under contract No.~DE-AC02-06CH11357$^1$,
and resources of the National Energy Research Scientific Computing Center, a
DOE Office of Science User Facility supported by the Office of Science of the
U.S. Department of Energy under contract No.~DE-AC02-05CH11231$^1$.  DIII-D
data shown in this paper can be obtained in digital format by following the
links at https://fusion.gat.com/global/D3D\_DMP.

\bibliographystyle{apsrev4-1}
\bibliography{Biblio}

\begin{thebibliography}{31}%
\makeatletter
\providecommand \@ifxundefined [1]{%
 \@ifx{#1\undefined}
}%
\providecommand \@ifnum [1]{%
 \ifnum #1\expandafter \@firstoftwo
 \else \expandafter \@secondoftwo
 \fi
}%
\providecommand \@ifx [1]{%
 \ifx #1\expandafter \@firstoftwo
 \else \expandafter \@secondoftwo
 \fi
}%
\providecommand \natexlab [1]{#1}%
\providecommand \enquote  [1]{``#1''}%
\providecommand \bibnamefont  [1]{#1}%
\providecommand \bibfnamefont [1]{#1}%
\providecommand \citenamefont [1]{#1}%
\providecommand \href@noop [0]{\@secondoftwo}%
\providecommand \href [0]{\begingroup \@sanitize@url \@href}%
\providecommand \@href[1]{\@@startlink{#1}\@@href}%
\providecommand \@@href[1]{\endgroup#1\@@endlink}%
\providecommand \@sanitize@url [0]{\catcode `\\12\catcode `\$12\catcode
  `\&12\catcode `\#12\catcode `\^12\catcode `\_12\catcode `\%12\relax}%
\providecommand \@@startlink[1]{}%
\providecommand \@@endlink[0]{}%
\providecommand \url  [0]{\begingroup\@sanitize@url \@url }%
\providecommand \@url [1]{\endgroup\@href {#1}{\urlprefix }}%
\providecommand \urlprefix  [0]{URL }%
\providecommand \Eprint [0]{\href }%
\providecommand \doibase [0]{http://dx.doi.org/}%
\providecommand \selectlanguage [0]{\@gobble}%
\providecommand \bibinfo  [0]{\@secondoftwo}%
\providecommand \bibfield  [0]{\@secondoftwo}%
\providecommand \translation [1]{[#1]}%
\providecommand \BibitemOpen [0]{}%
\providecommand \bibitemStop [0]{}%
\providecommand \bibitemNoStop [0]{.\EOS\space}%
\providecommand \EOS [0]{\spacefactor3000\relax}%
\providecommand \BibitemShut  [1]{\csname bibitem#1\endcsname}%
\let\auto@bib@innerbib\@empty
\bibitem [{\citenamefont {Garofalo}\ \emph {et~al.}(2015)\citenamefont
  {Garofalo}, \citenamefont {Burrell}, \citenamefont {Eldon}, \citenamefont
  {Grierson}, \citenamefont {Hanson}, \citenamefont {Holland}, \citenamefont
  {Huijsmans}, \citenamefont {Liu}, \citenamefont {Loarte}, \citenamefont
  {Meneghini}, \citenamefont {Osborne}, \citenamefont {Paz-Soldan},
  \citenamefont {Smith}, \citenamefont {Snyder}, \citenamefont {Solomon},
  \citenamefont {Turnbull},\ and\ \citenamefont {Zeng}}]{garofalo15}%
  \BibitemOpen
  \bibfield  {author} {\bibinfo {author} {\bibfnamefont {A.~M.}\ \bibnamefont
  {Garofalo}}, \bibinfo {author} {\bibfnamefont {K.~H.}\ \bibnamefont
  {Burrell}}, \bibinfo {author} {\bibfnamefont {D.}~\bibnamefont {Eldon}},
  \bibinfo {author} {\bibfnamefont {B.~A.}\ \bibnamefont {Grierson}}, \bibinfo
  {author} {\bibfnamefont {J.~M.}\ \bibnamefont {Hanson}}, \bibinfo {author}
  {\bibfnamefont {C.}~\bibnamefont {Holland}}, \bibinfo {author} {\bibfnamefont
  {G.~T.~A.}\ \bibnamefont {Huijsmans}}, \bibinfo {author} {\bibfnamefont
  {F.}~\bibnamefont {Liu}}, \bibinfo {author} {\bibfnamefont {A.}~\bibnamefont
  {Loarte}}, \bibinfo {author} {\bibfnamefont {O.}~\bibnamefont {Meneghini}},
  \bibinfo {author} {\bibfnamefont {T.~H.}\ \bibnamefont {Osborne}}, \bibinfo
  {author} {\bibfnamefont {C.}~\bibnamefont {Paz-Soldan}}, \bibinfo {author}
  {\bibfnamefont {S.~P.}\ \bibnamefont {Smith}}, \bibinfo {author}
  {\bibfnamefont {P.~B.}\ \bibnamefont {Snyder}}, \bibinfo {author}
  {\bibfnamefont {W.~M.}\ \bibnamefont {Solomon}}, \bibinfo {author}
  {\bibfnamefont {A.~D.}\ \bibnamefont {Turnbull}}, \ and\ \bibinfo {author}
  {\bibfnamefont {L.}~\bibnamefont {Zeng}},\ }\href {\doibase
  http://dx.doi.org/10.1063/1.4921406} {\bibfield  {journal} {\bibinfo
  {journal} {Physics of Plasmas}\ }\textbf {\bibinfo {volume} {22}},\ \bibinfo
  {pages} {056116} (\bibinfo {year} {2015})}\BibitemShut {NoStop}%
\bibitem [{\citenamefont {Connor}\ \emph {et~al.}(1998)\citenamefont {Connor},
  \citenamefont {Hastie}, \citenamefont {Wilson},\ and\ \citenamefont
  {Miller}}]{connor98}%
  \BibitemOpen
  \bibfield  {author} {\bibinfo {author} {\bibfnamefont {J.~W.}\ \bibnamefont
  {Connor}}, \bibinfo {author} {\bibfnamefont {R.~J.}\ \bibnamefont {Hastie}},
  \bibinfo {author} {\bibfnamefont {H.~R.}\ \bibnamefont {Wilson}}, \ and\
  \bibinfo {author} {\bibfnamefont {R.~L.}\ \bibnamefont {Miller}},\ }\href
  {\doibase http://dx.doi.org/10.1063/1.872956} {\bibfield  {journal} {\bibinfo
   {journal} {Physics of Plasmas}\ }\textbf {\bibinfo {volume} {5}},\ \bibinfo
  {pages} {2687} (\bibinfo {year} {1998})}\BibitemShut {NoStop}%
\bibitem [{\citenamefont {Leonard}\ \emph {et~al.}(2006)\citenamefont
  {Leonard}, \citenamefont {Asakura}, \citenamefont {Boedo}, \citenamefont
  {Becoulet}, \citenamefont {Counsell}, \citenamefont {Eich}, \citenamefont
  {Fundamenski}, \citenamefont {Herrmann}, \citenamefont {Horton},
  \citenamefont {Kamada}, \citenamefont {Kirk}, \citenamefont {Kurzan},
  \citenamefont {Loarte}, \citenamefont {Neuhauser}, \citenamefont {Nunes},
  \citenamefont {Oyama}, \citenamefont {Pitts}, \citenamefont {Saibene},
  \citenamefont {Silva}, \citenamefont {Snyder}, \citenamefont {Urano},
  \citenamefont {Wade}, \citenamefont {Wilson},\ and\ \citenamefont
  {Group}}]{leonard06}%
  \BibitemOpen
  \bibfield  {author} {\bibinfo {author} {\bibfnamefont {A.~W.}\ \bibnamefont
  {Leonard}}, \bibinfo {author} {\bibfnamefont {N.}~\bibnamefont {Asakura}},
  \bibinfo {author} {\bibfnamefont {J.~A.}\ \bibnamefont {Boedo}}, \bibinfo
  {author} {\bibfnamefont {M.}~\bibnamefont {Becoulet}}, \bibinfo {author}
  {\bibfnamefont {G.~F.}\ \bibnamefont {Counsell}}, \bibinfo {author}
  {\bibfnamefont {T.}~\bibnamefont {Eich}}, \bibinfo {author} {\bibfnamefont
  {W.}~\bibnamefont {Fundamenski}}, \bibinfo {author} {\bibfnamefont
  {A.}~\bibnamefont {Herrmann}}, \bibinfo {author} {\bibfnamefont {L.~D.}\
  \bibnamefont {Horton}}, \bibinfo {author} {\bibfnamefont {Y.}~\bibnamefont
  {Kamada}}, \bibinfo {author} {\bibfnamefont {A.}~\bibnamefont {Kirk}},
  \bibinfo {author} {\bibfnamefont {B.}~\bibnamefont {Kurzan}}, \bibinfo
  {author} {\bibfnamefont {A.}~\bibnamefont {Loarte}}, \bibinfo {author}
  {\bibfnamefont {J.}~\bibnamefont {Neuhauser}}, \bibinfo {author}
  {\bibfnamefont {I.}~\bibnamefont {Nunes}}, \bibinfo {author} {\bibfnamefont
  {N.}~\bibnamefont {Oyama}}, \bibinfo {author} {\bibfnamefont {R.~A.}\
  \bibnamefont {Pitts}}, \bibinfo {author} {\bibfnamefont {G.}~\bibnamefont
  {Saibene}}, \bibinfo {author} {\bibfnamefont {C.}~\bibnamefont {Silva}},
  \bibinfo {author} {\bibfnamefont {P.~B.}\ \bibnamefont {Snyder}}, \bibinfo
  {author} {\bibfnamefont {H.}~\bibnamefont {Urano}}, \bibinfo {author}
  {\bibfnamefont {M.~R.}\ \bibnamefont {Wade}}, \bibinfo {author}
  {\bibfnamefont {H.~R.}\ \bibnamefont {Wilson}}, \ and\ \bibinfo {author}
  {\bibfnamefont {f.~t. P. a. E. P.~I.}\ \bibnamefont {Group}},\ }\href
  {\doibase 10.1088/0741-3335/48/5A/S14} {\bibfield  {journal} {\bibinfo
  {journal} {Plasma Physics and Controlled Fusion}\ }\textbf {\bibinfo {volume}
  {48}},\ \bibinfo {pages} {A149} (\bibinfo {year} {2006})}\BibitemShut
  {NoStop}%
\bibitem [{\citenamefont {Burrell}\ \emph {et~al.}(2005)\citenamefont
  {Burrell}, \citenamefont {West}, \citenamefont {Doyle}, \citenamefont
  {Austin}, \citenamefont {Casper}, \citenamefont {Gohil}, \citenamefont
  {Greenfield}, \citenamefont {Groebner}, \citenamefont {Hyatt}, \citenamefont
  {Jayakumar}, \citenamefont {Kaplan}, \citenamefont {Lao}, \citenamefont
  {Leonard}, \citenamefont {Makowski}, \citenamefont {McKee}, \citenamefont
  {Osborne}, \citenamefont {Snyder}, \citenamefont {Solomon}, \citenamefont
  {Thomas}, \citenamefont {Rhodes}, \citenamefont {Strait}, \citenamefont
  {Wade}, \citenamefont {Wang},\ and\ \citenamefont {Zeng}}]{burrell05}%
  \BibitemOpen
  \bibfield  {author} {\bibinfo {author} {\bibfnamefont {K.~H.}\ \bibnamefont
  {Burrell}}, \bibinfo {author} {\bibfnamefont {W.~P.}\ \bibnamefont {West}},
  \bibinfo {author} {\bibfnamefont {E.~J.}\ \bibnamefont {Doyle}}, \bibinfo
  {author} {\bibfnamefont {M.~E.}\ \bibnamefont {Austin}}, \bibinfo {author}
  {\bibfnamefont {T.~A.}\ \bibnamefont {Casper}}, \bibinfo {author}
  {\bibfnamefont {P.}~\bibnamefont {Gohil}}, \bibinfo {author} {\bibfnamefont
  {C.~M.}\ \bibnamefont {Greenfield}}, \bibinfo {author} {\bibfnamefont
  {R.~J.}\ \bibnamefont {Groebner}}, \bibinfo {author} {\bibfnamefont {A.~W.}\
  \bibnamefont {Hyatt}}, \bibinfo {author} {\bibfnamefont {R.~J.}\ \bibnamefont
  {Jayakumar}}, \bibinfo {author} {\bibfnamefont {D.~H.}\ \bibnamefont
  {Kaplan}}, \bibinfo {author} {\bibfnamefont {L.~L.}\ \bibnamefont {Lao}},
  \bibinfo {author} {\bibfnamefont {A.~W.}\ \bibnamefont {Leonard}}, \bibinfo
  {author} {\bibfnamefont {M.~A.}\ \bibnamefont {Makowski}}, \bibinfo {author}
  {\bibfnamefont {G.~R.}\ \bibnamefont {McKee}}, \bibinfo {author}
  {\bibfnamefont {T.~H.}\ \bibnamefont {Osborne}}, \bibinfo {author}
  {\bibfnamefont {P.~B.}\ \bibnamefont {Snyder}}, \bibinfo {author}
  {\bibfnamefont {W.~M.}\ \bibnamefont {Solomon}}, \bibinfo {author}
  {\bibfnamefont {D.~M.}\ \bibnamefont {Thomas}}, \bibinfo {author}
  {\bibfnamefont {T.~L.}\ \bibnamefont {Rhodes}}, \bibinfo {author}
  {\bibfnamefont {E.~J.}\ \bibnamefont {Strait}}, \bibinfo {author}
  {\bibfnamefont {M.~R.}\ \bibnamefont {Wade}}, \bibinfo {author}
  {\bibfnamefont {G.}~\bibnamefont {Wang}}, \ and\ \bibinfo {author}
  {\bibfnamefont {L.}~\bibnamefont {Zeng}},\ }\href {\doibase
  http://dx.doi.org/10.1063/1.1894745} {\bibfield  {journal} {\bibinfo
  {journal} {Physics of Plasmas}\ }\textbf {\bibinfo {volume} {12}},\ \bibinfo
  {pages} {056121} (\bibinfo {year} {2005})}\BibitemShut {NoStop}%
\bibitem [{\citenamefont {Solomon}\ \emph {et~al.}(2014)\citenamefont
  {Solomon}, \citenamefont {Snyder}, \citenamefont {Burrell}, \citenamefont
  {Fenstermacher}, \citenamefont {Garofalo}, \citenamefont {Grierson},
  \citenamefont {Loarte}, \citenamefont {McKee}, \citenamefont {Nazikian},\
  and\ \citenamefont {Osborne}}]{solomon14}%
  \BibitemOpen
  \bibfield  {author} {\bibinfo {author} {\bibfnamefont {W.~M.}\ \bibnamefont
  {Solomon}}, \bibinfo {author} {\bibfnamefont {P.~B.}\ \bibnamefont {Snyder}},
  \bibinfo {author} {\bibfnamefont {K.~H.}\ \bibnamefont {Burrell}}, \bibinfo
  {author} {\bibfnamefont {M.~E.}\ \bibnamefont {Fenstermacher}}, \bibinfo
  {author} {\bibfnamefont {A.~M.}\ \bibnamefont {Garofalo}}, \bibinfo {author}
  {\bibfnamefont {B.~A.}\ \bibnamefont {Grierson}}, \bibinfo {author}
  {\bibfnamefont {A.}~\bibnamefont {Loarte}}, \bibinfo {author} {\bibfnamefont
  {G.~R.}\ \bibnamefont {McKee}}, \bibinfo {author} {\bibfnamefont
  {R.}~\bibnamefont {Nazikian}}, \ and\ \bibinfo {author} {\bibfnamefont
  {T.~H.}\ \bibnamefont {Osborne}},\ }\href {\doibase
  10.1103/PhysRevLett.113.135001} {\bibfield  {journal} {\bibinfo  {journal}
  {Phys. Rev. Lett.}\ }\textbf {\bibinfo {volume} {113}},\ \bibinfo {pages}
  {135001} (\bibinfo {year} {2014})}\BibitemShut {NoStop}%
\bibitem [{\citenamefont {Grierson}\ \emph {et~al.}(2015)\citenamefont
  {Grierson}, \citenamefont {Burrell}, \citenamefont {Nazikian}, \citenamefont
  {Solomon}, \citenamefont {Garofalo}, \citenamefont {Belli}, \citenamefont
  {Staebler}, \citenamefont {Fenstermacher}, \citenamefont {McKee},
  \citenamefont {Evans}, \citenamefont {Orlov}, \citenamefont {Smith},
  \citenamefont {Chrobak}, \citenamefont {Chrystal},\ and\ \citenamefont
  {Team}}]{grierson15}%
  \BibitemOpen
  \bibfield  {author} {\bibinfo {author} {\bibfnamefont {B.~A.}\ \bibnamefont
  {Grierson}}, \bibinfo {author} {\bibfnamefont {K.~H.}\ \bibnamefont
  {Burrell}}, \bibinfo {author} {\bibfnamefont {R.~M.}\ \bibnamefont
  {Nazikian}}, \bibinfo {author} {\bibfnamefont {W.~M.}\ \bibnamefont
  {Solomon}}, \bibinfo {author} {\bibfnamefont {A.~M.}\ \bibnamefont
  {Garofalo}}, \bibinfo {author} {\bibfnamefont {E.~A.}\ \bibnamefont {Belli}},
  \bibinfo {author} {\bibfnamefont {G.~M.}\ \bibnamefont {Staebler}}, \bibinfo
  {author} {\bibfnamefont {M.~E.}\ \bibnamefont {Fenstermacher}}, \bibinfo
  {author} {\bibfnamefont {G.~R.}\ \bibnamefont {McKee}}, \bibinfo {author}
  {\bibfnamefont {T.~E.}\ \bibnamefont {Evans}}, \bibinfo {author}
  {\bibfnamefont {D.~M.}\ \bibnamefont {Orlov}}, \bibinfo {author}
  {\bibfnamefont {S.~P.}\ \bibnamefont {Smith}}, \bibinfo {author}
  {\bibfnamefont {C.}~\bibnamefont {Chrobak}}, \bibinfo {author} {\bibfnamefont
  {C.}~\bibnamefont {Chrystal}}, \ and\ \bibinfo {author} {\bibfnamefont
  {D.-D.}\ \bibnamefont {Team}},\ }\href {\doibase
  http://dx.doi.org/10.1063/1.4918359} {\bibfield  {journal} {\bibinfo
  {journal} {Physics of Plasmas}\ }\textbf {\bibinfo {volume} {22}},\ \bibinfo
  {pages} {055901} (\bibinfo {year} {2015})}\BibitemShut {NoStop}%
\bibitem [{\citenamefont {Burrell}\ \emph {et~al.}(2016)\citenamefont
  {Burrell}, \citenamefont {Barada}, \citenamefont {Chen}, \citenamefont
  {Garofalo}, \citenamefont {Groebner}, \citenamefont {Muscatello},
  \citenamefont {Osborne}, \citenamefont {Petty}, \citenamefont {Rhodes},
  \citenamefont {Snyder}, \citenamefont {Solomon}, \citenamefont {Yan},\ and\
  \citenamefont {Zeng}}]{burrell16}%
  \BibitemOpen
  \bibfield  {author} {\bibinfo {author} {\bibfnamefont {K.~H.}\ \bibnamefont
  {Burrell}}, \bibinfo {author} {\bibfnamefont {K.}~\bibnamefont {Barada}},
  \bibinfo {author} {\bibfnamefont {X.}~\bibnamefont {Chen}}, \bibinfo {author}
  {\bibfnamefont {A.~M.}\ \bibnamefont {Garofalo}}, \bibinfo {author}
  {\bibfnamefont {R.~J.}\ \bibnamefont {Groebner}}, \bibinfo {author}
  {\bibfnamefont {C.~M.}\ \bibnamefont {Muscatello}}, \bibinfo {author}
  {\bibfnamefont {T.~H.}\ \bibnamefont {Osborne}}, \bibinfo {author}
  {\bibfnamefont {C.~C.}\ \bibnamefont {Petty}}, \bibinfo {author}
  {\bibfnamefont {T.~L.}\ \bibnamefont {Rhodes}}, \bibinfo {author}
  {\bibfnamefont {P.~B.}\ \bibnamefont {Snyder}}, \bibinfo {author}
  {\bibfnamefont {W.~M.}\ \bibnamefont {Solomon}}, \bibinfo {author}
  {\bibfnamefont {Z.}~\bibnamefont {Yan}}, \ and\ \bibinfo {author}
  {\bibfnamefont {L.}~\bibnamefont {Zeng}},\ }\href {\doibase
  http://dx.doi.org/10.1063/1.4943521} {\bibfield  {journal} {\bibinfo
  {journal} {Physics of Plasmas}\ }\textbf {\bibinfo {volume} {23}},\ \bibinfo
  {pages} {056103} (\bibinfo {year} {2016})}\BibitemShut {NoStop}%
\bibitem [{\citenamefont {Chen}\ \emph {et~al.}(2017)\citenamefont {Chen},
  \citenamefont {Burrell}, \citenamefont {Osborne}, \citenamefont {Solomon},
  \citenamefont {Barada}, \citenamefont {Garofalo}, \citenamefont {Groebner},
  \citenamefont {Luhmann}, \citenamefont {McKee}, \citenamefont {Muscatello},
  \citenamefont {Ono}, \citenamefont {Petty}, \citenamefont {Porkolab},
  \citenamefont {Rhodes}, \citenamefont {Rost}, \citenamefont {Snyder},
  \citenamefont {Staebler}, \citenamefont {Tobias}, \citenamefont {Yan},\ and\
  \citenamefont {the DIII-D~Team}}]{chen17}%
  \BibitemOpen
  \bibfield  {author} {\bibinfo {author} {\bibfnamefont {X.}~\bibnamefont
  {Chen}}, \bibinfo {author} {\bibfnamefont {K.}~\bibnamefont {Burrell}},
  \bibinfo {author} {\bibfnamefont {T.}~\bibnamefont {Osborne}}, \bibinfo
  {author} {\bibfnamefont {W.}~\bibnamefont {Solomon}}, \bibinfo {author}
  {\bibfnamefont {K.}~\bibnamefont {Barada}}, \bibinfo {author} {\bibfnamefont
  {A.}~\bibnamefont {Garofalo}}, \bibinfo {author} {\bibfnamefont
  {R.}~\bibnamefont {Groebner}}, \bibinfo {author} {\bibfnamefont
  {N.}~\bibnamefont {Luhmann}}, \bibinfo {author} {\bibfnamefont
  {G.}~\bibnamefont {McKee}}, \bibinfo {author} {\bibfnamefont
  {C.}~\bibnamefont {Muscatello}}, \bibinfo {author} {\bibfnamefont
  {M.}~\bibnamefont {Ono}}, \bibinfo {author} {\bibfnamefont {C.}~\bibnamefont
  {Petty}}, \bibinfo {author} {\bibfnamefont {M.}~\bibnamefont {Porkolab}},
  \bibinfo {author} {\bibfnamefont {T.}~\bibnamefont {Rhodes}}, \bibinfo
  {author} {\bibfnamefont {J.}~\bibnamefont {Rost}}, \bibinfo {author}
  {\bibfnamefont {P.}~\bibnamefont {Snyder}}, \bibinfo {author} {\bibfnamefont
  {G.}~\bibnamefont {Staebler}}, \bibinfo {author} {\bibfnamefont
  {B.}~\bibnamefont {Tobias}}, \bibinfo {author} {\bibfnamefont
  {Z.}~\bibnamefont {Yan}}, \ and\ \bibinfo {author} {\bibnamefont {the
  DIII-D~Team}},\ }\href {http://stacks.iop.org/0029-5515/57/i=2/a=022007}
  {\bibfield  {journal} {\bibinfo  {journal} {Nuclear Fusion}\ }\textbf
  {\bibinfo {volume} {57}},\ \bibinfo {pages} {022007} (\bibinfo {year}
  {2017})}\BibitemShut {NoStop}%
\bibitem [{\citenamefont {Sakamoto}\ \emph {et~al.}(2004)\citenamefont
  {Sakamoto}, \citenamefont {Shirai}, \citenamefont {Fujita}, \citenamefont
  {Ide}, \citenamefont {Takizuka}, \citenamefont {Oyama},\ and\ \citenamefont
  {Kamada}}]{Sakamoto04}%
  \BibitemOpen
  \bibfield  {author} {\bibinfo {author} {\bibfnamefont {Y.}~\bibnamefont
  {Sakamoto}}, \bibinfo {author} {\bibfnamefont {H.}~\bibnamefont {Shirai}},
  \bibinfo {author} {\bibfnamefont {T.}~\bibnamefont {Fujita}}, \bibinfo
  {author} {\bibfnamefont {S.}~\bibnamefont {Ide}}, \bibinfo {author}
  {\bibfnamefont {T.}~\bibnamefont {Takizuka}}, \bibinfo {author}
  {\bibfnamefont {N.}~\bibnamefont {Oyama}}, \ and\ \bibinfo {author}
  {\bibfnamefont {Y.}~\bibnamefont {Kamada}},\ }\href
  {http://stacks.iop.org/0741-3335/46/i=5A/a=033} {\bibfield  {journal}
  {\bibinfo  {journal} {Plasma Physics and Controlled Fusion}\ }\textbf
  {\bibinfo {volume} {46}},\ \bibinfo {pages} {A299} (\bibinfo {year}
  {2004})}\BibitemShut {NoStop}%
\bibitem [{\citenamefont {Oyama}\ \emph {et~al.}(2005)\citenamefont {Oyama},
  \citenamefont {Sakamoto}, \citenamefont {Isayama}, \citenamefont {Takechi},
  \citenamefont {Gohil}, \citenamefont {Lao}, \citenamefont {Snyder},
  \citenamefont {Fujita}, \citenamefont {Ide}, \citenamefont {Kamada},
  \citenamefont {Miura}, \citenamefont {Oikawa}, \citenamefont {Suzuki},
  \citenamefont {Takenaga}, \citenamefont {Toi},\ and\ \citenamefont {the
  JT-60~Team}}]{oyama05}%
  \BibitemOpen
  \bibfield  {author} {\bibinfo {author} {\bibfnamefont {N.}~\bibnamefont
  {Oyama}}, \bibinfo {author} {\bibfnamefont {Y.}~\bibnamefont {Sakamoto}},
  \bibinfo {author} {\bibfnamefont {A.}~\bibnamefont {Isayama}}, \bibinfo
  {author} {\bibfnamefont {M.}~\bibnamefont {Takechi}}, \bibinfo {author}
  {\bibfnamefont {P.}~\bibnamefont {Gohil}}, \bibinfo {author} {\bibfnamefont
  {L.}~\bibnamefont {Lao}}, \bibinfo {author} {\bibfnamefont {P.}~\bibnamefont
  {Snyder}}, \bibinfo {author} {\bibfnamefont {T.}~\bibnamefont {Fujita}},
  \bibinfo {author} {\bibfnamefont {S.}~\bibnamefont {Ide}}, \bibinfo {author}
  {\bibfnamefont {Y.}~\bibnamefont {Kamada}}, \bibinfo {author} {\bibfnamefont
  {Y.}~\bibnamefont {Miura}}, \bibinfo {author} {\bibfnamefont
  {T.}~\bibnamefont {Oikawa}}, \bibinfo {author} {\bibfnamefont
  {T.}~\bibnamefont {Suzuki}}, \bibinfo {author} {\bibfnamefont
  {H.}~\bibnamefont {Takenaga}}, \bibinfo {author} {\bibfnamefont
  {K.}~\bibnamefont {Toi}}, \ and\ \bibinfo {author} {\bibnamefont {the
  JT-60~Team}},\ }\href {http://stacks.iop.org/0029-5515/45/i=8/a=014}
  {\bibfield  {journal} {\bibinfo  {journal} {Nuclear Fusion}\ }\textbf
  {\bibinfo {volume} {45}},\ \bibinfo {pages} {871} (\bibinfo {year}
  {2005})}\BibitemShut {NoStop}%
\bibitem [{\citenamefont {Solano}\ \emph {et~al.}(2010)\citenamefont {Solano},
  \citenamefont {Lomas}, \citenamefont {Alper}, \citenamefont {Xu},
  \citenamefont {Andrew}, \citenamefont {Arnoux}, \citenamefont {Boboc},
  \citenamefont {Barrera}, \citenamefont {Belo}, \citenamefont {Beurskens},
  \citenamefont {Brix}, \citenamefont {Crombe}, \citenamefont {de~la Luna},
  \citenamefont {Devaux}, \citenamefont {Eich}, \citenamefont {Gerasimov},
  \citenamefont {Giroud}, \citenamefont {Harting}, \citenamefont {Howell},
  \citenamefont {Huber}, \citenamefont {Kocsis}, \citenamefont {Korotkov},
  \citenamefont {Lopez-Fraguas}, \citenamefont {Nave}, \citenamefont {Rachlew},
  \citenamefont {Rimini}, \citenamefont {Saarelma}, \citenamefont {Sirinelli},
  \citenamefont {Pinches}, \citenamefont {Thomsen}, \citenamefont {Zabeo},\
  and\ \citenamefont {Zarzoso}}]{solano10}%
  \BibitemOpen
  \bibfield  {author} {\bibinfo {author} {\bibfnamefont {E.~R.}\ \bibnamefont
  {Solano}}, \bibinfo {author} {\bibfnamefont {P.~J.}\ \bibnamefont {Lomas}},
  \bibinfo {author} {\bibfnamefont {B.}~\bibnamefont {Alper}}, \bibinfo
  {author} {\bibfnamefont {G.~S.}\ \bibnamefont {Xu}}, \bibinfo {author}
  {\bibfnamefont {Y.}~\bibnamefont {Andrew}}, \bibinfo {author} {\bibfnamefont
  {G.}~\bibnamefont {Arnoux}}, \bibinfo {author} {\bibfnamefont
  {A.}~\bibnamefont {Boboc}}, \bibinfo {author} {\bibfnamefont
  {L.}~\bibnamefont {Barrera}}, \bibinfo {author} {\bibfnamefont
  {P.}~\bibnamefont {Belo}}, \bibinfo {author} {\bibfnamefont {M.~N.~A.}\
  \bibnamefont {Beurskens}}, \bibinfo {author} {\bibfnamefont {M.}~\bibnamefont
  {Brix}}, \bibinfo {author} {\bibfnamefont {K.}~\bibnamefont {Crombe}},
  \bibinfo {author} {\bibfnamefont {E.}~\bibnamefont {de~la Luna}}, \bibinfo
  {author} {\bibfnamefont {S.}~\bibnamefont {Devaux}}, \bibinfo {author}
  {\bibfnamefont {T.}~\bibnamefont {Eich}}, \bibinfo {author} {\bibfnamefont
  {S.}~\bibnamefont {Gerasimov}}, \bibinfo {author} {\bibfnamefont
  {C.}~\bibnamefont {Giroud}}, \bibinfo {author} {\bibfnamefont
  {D.}~\bibnamefont {Harting}}, \bibinfo {author} {\bibfnamefont
  {D.}~\bibnamefont {Howell}}, \bibinfo {author} {\bibfnamefont
  {A.}~\bibnamefont {Huber}}, \bibinfo {author} {\bibfnamefont
  {G.}~\bibnamefont {Kocsis}}, \bibinfo {author} {\bibfnamefont
  {A.}~\bibnamefont {Korotkov}}, \bibinfo {author} {\bibfnamefont
  {A.}~\bibnamefont {Lopez-Fraguas}}, \bibinfo {author} {\bibfnamefont
  {M.~F.~F.}\ \bibnamefont {Nave}}, \bibinfo {author} {\bibfnamefont
  {E.}~\bibnamefont {Rachlew}}, \bibinfo {author} {\bibfnamefont
  {F.}~\bibnamefont {Rimini}}, \bibinfo {author} {\bibfnamefont
  {S.}~\bibnamefont {Saarelma}}, \bibinfo {author} {\bibfnamefont
  {A.}~\bibnamefont {Sirinelli}}, \bibinfo {author} {\bibfnamefont {S.~D.}\
  \bibnamefont {Pinches}}, \bibinfo {author} {\bibfnamefont {H.}~\bibnamefont
  {Thomsen}}, \bibinfo {author} {\bibfnamefont {L.}~\bibnamefont {Zabeo}}, \
  and\ \bibinfo {author} {\bibfnamefont {D.}~\bibnamefont {Zarzoso}},\ }\href
  {\doibase 10.1103/PhysRevLett.104.185003} {\bibfield  {journal} {\bibinfo
  {journal} {Phys. Rev. Lett.}\ }\textbf {\bibinfo {volume} {104}},\ \bibinfo
  {pages} {185003} (\bibinfo {year} {2010})}\BibitemShut {NoStop}%
\bibitem [{\citenamefont {Suttrop}\ \emph {et~al.}(2005)\citenamefont
  {Suttrop}, \citenamefont {Hynönen}, \citenamefont {Kurki-Suonio},
  \citenamefont {Lang}, \citenamefont {Maraschek}, \citenamefont {Neu},
  \citenamefont {Stäbler}, \citenamefont {Conway}, \citenamefont {Hacquin},
  \citenamefont {Kempenaars}, \citenamefont {Lomas}, \citenamefont {Nave},
  \citenamefont {Pitts}, \citenamefont {Zastrow}, \citenamefont {the ASDEX
  Upgrade~team},\ and\ \citenamefont {contributors to~the
  JET-EFDA~workprogramme}}]{suttrop05}%
  \BibitemOpen
  \bibfield  {author} {\bibinfo {author} {\bibfnamefont {W.}~\bibnamefont
  {Suttrop}}, \bibinfo {author} {\bibfnamefont {V.}~\bibnamefont {Hynönen}},
  \bibinfo {author} {\bibfnamefont {T.}~\bibnamefont {Kurki-Suonio}}, \bibinfo
  {author} {\bibfnamefont {P.}~\bibnamefont {Lang}}, \bibinfo {author}
  {\bibfnamefont {M.}~\bibnamefont {Maraschek}}, \bibinfo {author}
  {\bibfnamefont {R.}~\bibnamefont {Neu}}, \bibinfo {author} {\bibfnamefont
  {A.}~\bibnamefont {Stäbler}}, \bibinfo {author} {\bibfnamefont
  {G.}~\bibnamefont {Conway}}, \bibinfo {author} {\bibfnamefont
  {S.}~\bibnamefont {Hacquin}}, \bibinfo {author} {\bibfnamefont
  {M.}~\bibnamefont {Kempenaars}}, \bibinfo {author} {\bibfnamefont
  {P.}~\bibnamefont {Lomas}}, \bibinfo {author} {\bibfnamefont
  {M.}~\bibnamefont {Nave}}, \bibinfo {author} {\bibfnamefont {R.}~\bibnamefont
  {Pitts}}, \bibinfo {author} {\bibfnamefont {K.-D.}\ \bibnamefont {Zastrow}},
  \bibinfo {author} {\bibnamefont {the ASDEX Upgrade~team}}, \ and\ \bibinfo
  {author} {\bibnamefont {contributors to~the JET-EFDA~workprogramme}},\ }\href
  {http://stacks.iop.org/0029-5515/45/i=7/a=021} {\bibfield  {journal}
  {\bibinfo  {journal} {Nuclear Fusion}\ }\textbf {\bibinfo {volume} {45}},\
  \bibinfo {pages} {721} (\bibinfo {year} {2005})}\BibitemShut {NoStop}%
\bibitem [{\citenamefont {Snyder}\ \emph {et~al.}(2007)\citenamefont {Snyder},
  \citenamefont {Burrell}, \citenamefont {Wilson}, \citenamefont {Chu},
  \citenamefont {Fenstermacher}, \citenamefont {Leonard}, \citenamefont
  {Moyer}, \citenamefont {Osborne}, \citenamefont {Umansky}, \citenamefont
  {West},\ and\ \citenamefont {Xu}}]{Snyder07}%
  \BibitemOpen
  \bibfield  {author} {\bibinfo {author} {\bibfnamefont {P.}~\bibnamefont
  {Snyder}}, \bibinfo {author} {\bibfnamefont {K.}~\bibnamefont {Burrell}},
  \bibinfo {author} {\bibfnamefont {H.}~\bibnamefont {Wilson}}, \bibinfo
  {author} {\bibfnamefont {M.}~\bibnamefont {Chu}}, \bibinfo {author}
  {\bibfnamefont {M.}~\bibnamefont {Fenstermacher}}, \bibinfo {author}
  {\bibfnamefont {A.}~\bibnamefont {Leonard}}, \bibinfo {author} {\bibfnamefont
  {R.}~\bibnamefont {Moyer}}, \bibinfo {author} {\bibfnamefont
  {T.}~\bibnamefont {Osborne}}, \bibinfo {author} {\bibfnamefont
  {M.}~\bibnamefont {Umansky}}, \bibinfo {author} {\bibfnamefont
  {W.}~\bibnamefont {West}}, \ and\ \bibinfo {author} {\bibfnamefont
  {X.}~\bibnamefont {Xu}},\ }\href
  {http://stacks.iop.org/0029-5515/47/i=8/a=030} {\bibfield  {journal}
  {\bibinfo  {journal} {Nuclear Fusion}\ }\textbf {\bibinfo {volume} {47}},\
  \bibinfo {pages} {961} (\bibinfo {year} {2007})}\BibitemShut {NoStop}%
\bibitem [{\citenamefont {Chen}\ \emph {et~al.}(2016)\citenamefont {Chen},
  \citenamefont {Burrell}, \citenamefont {Ferraro}, \citenamefont {Osborne},
  \citenamefont {Austin}, \citenamefont {Garofalo}, \citenamefont {Groebner},
  \citenamefont {Kramer}, \citenamefont {Jr}, \citenamefont {McKee},
  \citenamefont {Muscatello}, \citenamefont {Nazikian}, \citenamefont {Ren},
  \citenamefont {Snyder}, \citenamefont {Solomon}, \citenamefont {Tobias},\
  and\ \citenamefont {Yan}}]{chen16}%
  \BibitemOpen
  \bibfield  {author} {\bibinfo {author} {\bibfnamefont {X.}~\bibnamefont
  {Chen}}, \bibinfo {author} {\bibfnamefont {K.}~\bibnamefont {Burrell}},
  \bibinfo {author} {\bibfnamefont {N.}~\bibnamefont {Ferraro}}, \bibinfo
  {author} {\bibfnamefont {T.}~\bibnamefont {Osborne}}, \bibinfo {author}
  {\bibfnamefont {M.}~\bibnamefont {Austin}}, \bibinfo {author} {\bibfnamefont
  {A.}~\bibnamefont {Garofalo}}, \bibinfo {author} {\bibfnamefont
  {R.}~\bibnamefont {Groebner}}, \bibinfo {author} {\bibfnamefont
  {G.}~\bibnamefont {Kramer}}, \bibinfo {author} {\bibfnamefont {N.~L.}\
  \bibnamefont {Jr}}, \bibinfo {author} {\bibfnamefont {G.}~\bibnamefont
  {McKee}}, \bibinfo {author} {\bibfnamefont {C.}~\bibnamefont {Muscatello}},
  \bibinfo {author} {\bibfnamefont {R.}~\bibnamefont {Nazikian}}, \bibinfo
  {author} {\bibfnamefont {X.}~\bibnamefont {Ren}}, \bibinfo {author}
  {\bibfnamefont {P.}~\bibnamefont {Snyder}}, \bibinfo {author} {\bibfnamefont
  {W.}~\bibnamefont {Solomon}}, \bibinfo {author} {\bibfnamefont
  {B.}~\bibnamefont {Tobias}}, \ and\ \bibinfo {author} {\bibfnamefont
  {Z.}~\bibnamefont {Yan}},\ }\href
  {http://stacks.iop.org/0029-5515/56/i=7/a=076011} {\bibfield  {journal}
  {\bibinfo  {journal} {Nuclear Fusion}\ }\textbf {\bibinfo {volume} {56}},\
  \bibinfo {pages} {076011} (\bibinfo {year} {2016})}\BibitemShut {NoStop}%
\bibitem [{\citenamefont {Liu}\ \emph {et~al.}(2015)\citenamefont {Liu},
  \citenamefont {Huijsmans}, \citenamefont {Loarte}, \citenamefont {Garofalo},
  \citenamefont {Solomon}, \citenamefont {Snyder}, \citenamefont {Hoelzl},\
  and\ \citenamefont {Zeng}}]{liu15}%
  \BibitemOpen
  \bibfield  {author} {\bibinfo {author} {\bibfnamefont {F.}~\bibnamefont
  {Liu}}, \bibinfo {author} {\bibfnamefont {G.}~\bibnamefont {Huijsmans}},
  \bibinfo {author} {\bibfnamefont {A.}~\bibnamefont {Loarte}}, \bibinfo
  {author} {\bibfnamefont {A.}~\bibnamefont {Garofalo}}, \bibinfo {author}
  {\bibfnamefont {W.}~\bibnamefont {Solomon}}, \bibinfo {author} {\bibfnamefont
  {P.}~\bibnamefont {Snyder}}, \bibinfo {author} {\bibfnamefont
  {M.}~\bibnamefont {Hoelzl}}, \ and\ \bibinfo {author} {\bibfnamefont
  {L.}~\bibnamefont {Zeng}},\ }\href
  {http://stacks.iop.org/0029-5515/55/i=11/a=113002} {\bibfield  {journal}
  {\bibinfo  {journal} {Nuclear Fusion}\ }\textbf {\bibinfo {volume} {55}},\
  \bibinfo {pages} {113002} (\bibinfo {year} {2015})}\BibitemShut {NoStop}%
\bibitem [{\citenamefont {King}\ \emph
  {et~al.}(2017{\natexlab{a}})\citenamefont {King}, \citenamefont {Burrell},
  \citenamefont {Garofalo}, \citenamefont {Groebner}, \citenamefont {Kruger},
  \citenamefont {Pankin},\ and\ \citenamefont {Snyder}}]{King17}%
  \BibitemOpen
  \bibfield  {author} {\bibinfo {author} {\bibfnamefont {J.}~\bibnamefont
  {King}}, \bibinfo {author} {\bibfnamefont {K.}~\bibnamefont {Burrell}},
  \bibinfo {author} {\bibfnamefont {A.}~\bibnamefont {Garofalo}}, \bibinfo
  {author} {\bibfnamefont {R.}~\bibnamefont {Groebner}}, \bibinfo {author}
  {\bibfnamefont {S.}~\bibnamefont {Kruger}}, \bibinfo {author} {\bibfnamefont
  {A.}~\bibnamefont {Pankin}}, \ and\ \bibinfo {author} {\bibfnamefont
  {P.}~\bibnamefont {Snyder}},\ }\href
  {http://stacks.iop.org/0029-5515/57/i=2/a=022002} {\bibfield  {journal}
  {\bibinfo  {journal} {Nuclear Fusion}\ }\textbf {\bibinfo {volume} {57}},\
  \bibinfo {pages} {022002} (\bibinfo {year} {2017}{\natexlab{a}})}\BibitemShut
  {NoStop}%
\bibitem [{\citenamefont {Sovinec}\ \emph {et~al.}(2004)\citenamefont
  {Sovinec}, \citenamefont {Glasser}, \citenamefont {Gianakon}, \citenamefont
  {Barnes}, \citenamefont {Nebel}, \citenamefont {Kruger}, \citenamefont
  {Schnack}, \citenamefont {Plimpton}, \citenamefont {Tarditi},\ and\
  \citenamefont {Chu}}]{Sovinec04}%
  \BibitemOpen
  \bibfield  {author} {\bibinfo {author} {\bibfnamefont {C.}~\bibnamefont
  {Sovinec}}, \bibinfo {author} {\bibfnamefont {A.}~\bibnamefont {Glasser}},
  \bibinfo {author} {\bibfnamefont {T.}~\bibnamefont {Gianakon}}, \bibinfo
  {author} {\bibfnamefont {D.}~\bibnamefont {Barnes}}, \bibinfo {author}
  {\bibfnamefont {R.}~\bibnamefont {Nebel}}, \bibinfo {author} {\bibfnamefont
  {S.}~\bibnamefont {Kruger}}, \bibinfo {author} {\bibfnamefont
  {D.}~\bibnamefont {Schnack}}, \bibinfo {author} {\bibfnamefont
  {S.}~\bibnamefont {Plimpton}}, \bibinfo {author} {\bibfnamefont
  {A.}~\bibnamefont {Tarditi}}, \ and\ \bibinfo {author} {\bibfnamefont
  {M.}~\bibnamefont {Chu}},\ }\href {\doibase
  http://dx.doi.org/10.1016/j.jcp.2003.10.004} {\bibfield  {journal} {\bibinfo
  {journal} {Journal of Computational Physics}\ }\textbf {\bibinfo {volume}
  {195}},\ \bibinfo {pages} {355 } (\bibinfo {year} {2004})}\BibitemShut
  {NoStop}%
\bibitem [{\citenamefont {Lao}\ \emph {et~al.}(1985)\citenamefont {Lao},
  \citenamefont {John}, \citenamefont {Stambaugh}, \citenamefont {Kellman},\
  and\ \citenamefont {Pfeiffer}}]{lao85}%
  \BibitemOpen
  \bibfield  {author} {\bibinfo {author} {\bibfnamefont {L.~L.}\ \bibnamefont
  {Lao}}, \bibinfo {author} {\bibfnamefont {H.~S.}\ \bibnamefont {John}},
  \bibinfo {author} {\bibfnamefont {R.~D.}\ \bibnamefont {Stambaugh}}, \bibinfo
  {author} {\bibfnamefont {A.~G.}\ \bibnamefont {Kellman}}, \ and\ \bibinfo
  {author} {\bibfnamefont {W.}~\bibnamefont {Pfeiffer}},\ }\href@noop {}
  {\bibfield  {journal} {\bibinfo  {journal} {Nuclear Fusion}\ }\textbf
  {\bibinfo {volume} {25}},\ \bibinfo {pages} {1611} (\bibinfo {year}
  {1985})}\BibitemShut {NoStop}%
\bibitem [{\citenamefont {Howell}\ and\ \citenamefont
  {Sovinec}(2014)}]{Howell14}%
  \BibitemOpen
  \bibfield  {author} {\bibinfo {author} {\bibfnamefont {E.}~\bibnamefont
  {Howell}}\ and\ \bibinfo {author} {\bibfnamefont {C.}~\bibnamefont
  {Sovinec}},\ }\href {\doibase http://dx.doi.org/10.1016/j.cpc.2014.02.008}
  {\bibfield  {journal} {\bibinfo  {journal} {Computer Physics Communications}\
  }\textbf {\bibinfo {volume} {185}},\ \bibinfo {pages} {1415 } (\bibinfo
  {year} {2014})}\BibitemShut {NoStop}%
\bibitem [{\citenamefont {King}\ \emph
  {et~al.}(2017{\natexlab{b}})\citenamefont {King}, \citenamefont {Kruger},
  \citenamefont {Groebner}, \citenamefont {Hanson}, \citenamefont {Hebert},
  \citenamefont {Held},\ and\ \citenamefont {Jepson}}]{King16sol}%
  \BibitemOpen
  \bibfield  {author} {\bibinfo {author} {\bibfnamefont {J.~R.}\ \bibnamefont
  {King}}, \bibinfo {author} {\bibfnamefont {S.~E.}\ \bibnamefont {Kruger}},
  \bibinfo {author} {\bibfnamefont {R.~J.}\ \bibnamefont {Groebner}}, \bibinfo
  {author} {\bibfnamefont {J.~D.}\ \bibnamefont {Hanson}}, \bibinfo {author}
  {\bibfnamefont {J.~D.}\ \bibnamefont {Hebert}}, \bibinfo {author}
  {\bibfnamefont {E.~D.}\ \bibnamefont {Held}}, \ and\ \bibinfo {author}
  {\bibfnamefont {J.~R.}\ \bibnamefont {Jepson}},\ }\href {\doibase
  10.1063/1.4972822} {\bibfield  {journal} {\bibinfo  {journal} {Physics of
  Plasmas}\ }\textbf {\bibinfo {volume} {24}},\ \bibinfo {pages} {012504}
  (\bibinfo {year} {2017}{\natexlab{b}})},\ \Eprint
  {http://arxiv.org/abs/http://dx.doi.org/10.1063/1.4972822}
  {http://dx.doi.org/10.1063/1.4972822} \BibitemShut {NoStop}%
\bibitem [{\citenamefont {Holland}\ \emph {et~al.}(2011)\citenamefont
  {Holland}, \citenamefont {Schmitz}, \citenamefont {Rhodes}, \citenamefont
  {Peebles}, \citenamefont {Hillesheim}, \citenamefont {Wang}, \citenamefont
  {Zeng}, \citenamefont {Doyle}, \citenamefont {Smith}, \citenamefont {Prater},
  \citenamefont {Burrell}, \citenamefont {Candy}, \citenamefont {Waltz},
  \citenamefont {Kinsey}, \citenamefont {Staebler}, \citenamefont {DeBoo},
  \citenamefont {Petty}, \citenamefont {McKee}, \citenamefont {Yan},\ and\
  \citenamefont {White}}]{Holland:2011ef}%
  \BibitemOpen
  \bibfield  {author} {\bibinfo {author} {\bibfnamefont {C.}~\bibnamefont
  {Holland}}, \bibinfo {author} {\bibfnamefont {L.}~\bibnamefont {Schmitz}},
  \bibinfo {author} {\bibfnamefont {T.~L.}\ \bibnamefont {Rhodes}}, \bibinfo
  {author} {\bibfnamefont {W.~A.}\ \bibnamefont {Peebles}}, \bibinfo {author}
  {\bibfnamefont {J.~C.}\ \bibnamefont {Hillesheim}}, \bibinfo {author}
  {\bibfnamefont {G.}~\bibnamefont {Wang}}, \bibinfo {author} {\bibfnamefont
  {L.}~\bibnamefont {Zeng}}, \bibinfo {author} {\bibfnamefont {E.~J.}\
  \bibnamefont {Doyle}}, \bibinfo {author} {\bibfnamefont {S.~P.}\ \bibnamefont
  {Smith}}, \bibinfo {author} {\bibfnamefont {R.}~\bibnamefont {Prater}},
  \bibinfo {author} {\bibfnamefont {K.~H.}\ \bibnamefont {Burrell}}, \bibinfo
  {author} {\bibfnamefont {J.}~\bibnamefont {Candy}}, \bibinfo {author}
  {\bibfnamefont {R.~E.}\ \bibnamefont {Waltz}}, \bibinfo {author}
  {\bibfnamefont {J.~E.}\ \bibnamefont {Kinsey}}, \bibinfo {author}
  {\bibfnamefont {G.~M.}\ \bibnamefont {Staebler}}, \bibinfo {author}
  {\bibfnamefont {J.~C.}\ \bibnamefont {DeBoo}}, \bibinfo {author}
  {\bibfnamefont {C.~C.}\ \bibnamefont {Petty}}, \bibinfo {author}
  {\bibfnamefont {G.~R.}\ \bibnamefont {McKee}}, \bibinfo {author}
  {\bibfnamefont {Z.}~\bibnamefont {Yan}}, \ and\ \bibinfo {author}
  {\bibfnamefont {A.~E.}\ \bibnamefont {White}},\ }\href {\doibase
  http://dx.doi.org/10.1063/1.3574518} {\bibfield  {journal} {\bibinfo
  {journal} {Physics of Plasmas}\ }\textbf {\bibinfo {volume} {18}},\ \bibinfo
  {pages} {056113} (\bibinfo {year} {2011})}\BibitemShut {NoStop}%
\bibitem [{\citenamefont {Garofalo}\ \emph {et~al.}(2011)\citenamefont
  {Garofalo}, \citenamefont {Solomon}, \citenamefont {Park}, \citenamefont
  {Burrell}, \citenamefont {DeBoo}, \citenamefont {Lanctot}, \citenamefont
  {McKee}, \citenamefont {Reimerdes}, \citenamefont {Schmitz}, \citenamefont
  {Schaffer},\ and\ \citenamefont {Snyder}}]{garofalo11}%
  \BibitemOpen
  \bibfield  {author} {\bibinfo {author} {\bibfnamefont {A.}~\bibnamefont
  {Garofalo}}, \bibinfo {author} {\bibfnamefont {W.}~\bibnamefont {Solomon}},
  \bibinfo {author} {\bibfnamefont {J.-K.}\ \bibnamefont {Park}}, \bibinfo
  {author} {\bibfnamefont {K.}~\bibnamefont {Burrell}}, \bibinfo {author}
  {\bibfnamefont {J.}~\bibnamefont {DeBoo}}, \bibinfo {author} {\bibfnamefont
  {M.}~\bibnamefont {Lanctot}}, \bibinfo {author} {\bibfnamefont
  {G.}~\bibnamefont {McKee}}, \bibinfo {author} {\bibfnamefont
  {H.}~\bibnamefont {Reimerdes}}, \bibinfo {author} {\bibfnamefont
  {L.}~\bibnamefont {Schmitz}}, \bibinfo {author} {\bibfnamefont
  {M.}~\bibnamefont {Schaffer}}, \ and\ \bibinfo {author} {\bibfnamefont
  {P.}~\bibnamefont {Snyder}},\ }\href
  {http://stacks.iop.org/0029-5515/51/i=8/a=083018} {\bibfield  {journal}
  {\bibinfo  {journal} {Nuclear Fusion}\ }\textbf {\bibinfo {volume} {51}},\
  \bibinfo {pages} {083018} (\bibinfo {year} {2011})}\BibitemShut {NoStop}%
\bibitem [{\citenamefont {Pankin}\ \emph {et~al.}(2007)\citenamefont {Pankin},
  \citenamefont {Bateman}, \citenamefont {Brennan}, \citenamefont {Kritz},
  \citenamefont {Kruger}, \citenamefont {Snyder}, \citenamefont {Sovinec},\
  and\ \citenamefont {the NIMROD~team}}]{pankin07}%
  \BibitemOpen
  \bibfield  {author} {\bibinfo {author} {\bibfnamefont {A.~Y.}\ \bibnamefont
  {Pankin}}, \bibinfo {author} {\bibfnamefont {G.}~\bibnamefont {Bateman}},
  \bibinfo {author} {\bibfnamefont {D.~P.}\ \bibnamefont {Brennan}}, \bibinfo
  {author} {\bibfnamefont {A.~H.}\ \bibnamefont {Kritz}}, \bibinfo {author}
  {\bibfnamefont {S.}~\bibnamefont {Kruger}}, \bibinfo {author} {\bibfnamefont
  {P.~B.}\ \bibnamefont {Snyder}}, \bibinfo {author} {\bibfnamefont
  {C.}~\bibnamefont {Sovinec}}, \ and\ \bibinfo {author} {\bibnamefont {the
  NIMROD~team}},\ }\href {http://stacks.iop.org/0741-3335/49/i=7/a=S04}
  {\bibfield  {journal} {\bibinfo  {journal} {Plasma Physics and Controlled
  Fusion}\ }\textbf {\bibinfo {volume} {49}},\ \bibinfo {pages} {S63} (\bibinfo
  {year} {2007})}\BibitemShut {NoStop}%
\bibitem [{\citenamefont {Sugiyama}\ and\ \citenamefont
  {Strauss}(2010)}]{sugiyama10}%
  \BibitemOpen
  \bibfield  {author} {\bibinfo {author} {\bibfnamefont {L.~E.}\ \bibnamefont
  {Sugiyama}}\ and\ \bibinfo {author} {\bibfnamefont {H.~R.}\ \bibnamefont
  {Strauss}},\ }\href {\doibase http://dx.doi.org/10.1063/1.3449301} {\bibfield
   {journal} {\bibinfo  {journal} {Physics of Plasmas}\ }\textbf {\bibinfo
  {volume} {17}},\ \bibinfo {pages} {062505} (\bibinfo {year}
  {2010})}\BibitemShut {NoStop}%
\bibitem [{\citenamefont {Huijsmans}\ and\ \citenamefont
  {Loarte}(2013)}]{huijsmans13}%
  \BibitemOpen
  \bibfield  {author} {\bibinfo {author} {\bibfnamefont {G.}~\bibnamefont
  {Huijsmans}}\ and\ \bibinfo {author} {\bibfnamefont {A.}~\bibnamefont
  {Loarte}},\ }\href {http://stacks.iop.org/0029-5515/53/i=12/a=123023}
  {\bibfield  {journal} {\bibinfo  {journal} {Nuclear Fusion}\ }\textbf
  {\bibinfo {volume} {53}},\ \bibinfo {pages} {123023} (\bibinfo {year}
  {2013})}\BibitemShut {NoStop}%
\bibitem [{\citenamefont {Guo}\ and\ \citenamefont {Diamond}(2015)}]{Guo15}%
  \BibitemOpen
  \bibfield  {author} {\bibinfo {author} {\bibfnamefont {Z.~B.}\ \bibnamefont
  {Guo}}\ and\ \bibinfo {author} {\bibfnamefont {P.~H.}\ \bibnamefont
  {Diamond}},\ }\href {\doibase 10.1103/PhysRevLett.114.145002} {\bibfield
  {journal} {\bibinfo  {journal} {Phys. Rev. Lett.}\ }\textbf {\bibinfo
  {volume} {114}},\ \bibinfo {pages} {145002} (\bibinfo {year}
  {2015})}\BibitemShut {NoStop}%
\bibitem [{\citenamefont {Coppi}(1964)}]{Coppi64}%
  \BibitemOpen
  \bibfield  {author} {\bibinfo {author} {\bibfnamefont {B.}~\bibnamefont
  {Coppi}},\ }\href {\doibase DOI:10.1063/1.1711405} {\bibfield  {journal}
  {\bibinfo  {journal} {Physics of Fluids}\ }\textbf {\bibinfo {volume} {7}},\
  \bibinfo {pages} {1501} (\bibinfo {year} {1964})}\BibitemShut {NoStop}%
\bibitem [{\citenamefont {King}\ and\ \citenamefont {Kruger}(2014)}]{King14}%
  \BibitemOpen
  \bibfield  {author} {\bibinfo {author} {\bibfnamefont {J.~R.}\ \bibnamefont
  {King}}\ and\ \bibinfo {author} {\bibfnamefont {S.~E.}\ \bibnamefont
  {Kruger}},\ }\href {\doibase 10.1063/1.4899036} {\bibfield  {journal}
  {\bibinfo  {journal} {Physics of Plasmas}\ }\textbf {\bibinfo {volume}
  {21}},\ \bibinfo {pages} {102113} (\bibinfo {year} {2014})}\BibitemShut
  {NoStop}%
\bibitem [{\citenamefont {King}\ \emph {et~al.}(2016)\citenamefont {King},
  \citenamefont {Pankin}, \citenamefont {Kruger},\ and\ \citenamefont
  {Snyder}}]{King16meudas}%
  \BibitemOpen
  \bibfield  {author} {\bibinfo {author} {\bibfnamefont {J.~R.}\ \bibnamefont
  {King}}, \bibinfo {author} {\bibfnamefont {A.~Y.}\ \bibnamefont {Pankin}},
  \bibinfo {author} {\bibfnamefont {S.~E.}\ \bibnamefont {Kruger}}, \ and\
  \bibinfo {author} {\bibfnamefont {P.~B.}\ \bibnamefont {Snyder}},\ }\href
  {\doibase http://dx.doi.org/10.1063/1.4954302} {\bibfield  {journal}
  {\bibinfo  {journal} {Physics of Plasmas}\ }\textbf {\bibinfo {volume}
  {23}},\ \bibinfo {eid} {062123} (\bibinfo {year} {2016})}\BibitemShut
  {NoStop}%
\bibitem [{\citenamefont {Terry}\ \emph {et~al.}(2008)\citenamefont {Terry},
  \citenamefont {Greenwald}, \citenamefont {Leboeuf}, \citenamefont {McKee},
  \citenamefont {Mikkelsen}, \citenamefont {Nevins}, \citenamefont {Newman},
  \citenamefont {Stotler}, \citenamefont {on~Verification}, \citenamefont
  {Validation}, \citenamefont {Organization},\ and\ \citenamefont
  {Force}}]{terry2008validation}%
  \BibitemOpen
  \bibfield  {author} {\bibinfo {author} {\bibfnamefont {P.~W.}\ \bibnamefont
  {Terry}}, \bibinfo {author} {\bibfnamefont {M.}~\bibnamefont {Greenwald}},
  \bibinfo {author} {\bibfnamefont {J.-N.}\ \bibnamefont {Leboeuf}}, \bibinfo
  {author} {\bibfnamefont {G.~R.}\ \bibnamefont {McKee}}, \bibinfo {author}
  {\bibfnamefont {D.~R.}\ \bibnamefont {Mikkelsen}}, \bibinfo {author}
  {\bibfnamefont {W.~M.}\ \bibnamefont {Nevins}}, \bibinfo {author}
  {\bibfnamefont {D.~E.}\ \bibnamefont {Newman}}, \bibinfo {author}
  {\bibfnamefont {D.~P.}\ \bibnamefont {Stotler}}, \bibinfo {author}
  {\bibfnamefont {T.~G.}\ \bibnamefont {on~Verification}}, \bibinfo {author}
  {\bibnamefont {Validation}}, \bibinfo {author} {\bibfnamefont {U.~B.~P.}\
  \bibnamefont {Organization}}, \ and\ \bibinfo {author} {\bibfnamefont
  {U.~T.~T.}\ \bibnamefont {Force}},\ }\href {\doibase
  http://dx.doi.org/10.1063/1.2928909} {\bibfield  {journal} {\bibinfo
  {journal} {Physics of Plasmas}\ }\textbf {\bibinfo {volume} {15}},\ \bibinfo
  {pages} {062503} (\bibinfo {year} {2008})}\BibitemShut {NoStop}%
\bibitem [{\citenamefont {Callen}\ \emph {et~al.}(2010)\citenamefont {Callen},
  \citenamefont {Groebner}, \citenamefont {Osborne}, \citenamefont {Canik},
  \citenamefont {Owen.}, \citenamefont {Pankin}, \citenamefont {Rafiq},
  \citenamefont {Rognlien},\ and\ \citenamefont {Stacey}}]{callen10}%
  \BibitemOpen
  \bibfield  {author} {\bibinfo {author} {\bibfnamefont {J.}~\bibnamefont
  {Callen}}, \bibinfo {author} {\bibfnamefont {R.}~\bibnamefont {Groebner}},
  \bibinfo {author} {\bibfnamefont {T.}~\bibnamefont {Osborne}}, \bibinfo
  {author} {\bibfnamefont {J.}~\bibnamefont {Canik}}, \bibinfo {author}
  {\bibfnamefont {L.}~\bibnamefont {Owen.}}, \bibinfo {author} {\bibfnamefont
  {A.}~\bibnamefont {Pankin}}, \bibinfo {author} {\bibfnamefont
  {T.}~\bibnamefont {Rafiq}}, \bibinfo {author} {\bibfnamefont
  {T.}~\bibnamefont {Rognlien}}, \ and\ \bibinfo {author} {\bibfnamefont
  {W.}~\bibnamefont {Stacey}},\ }\href
  {http://stacks.iop.org/0029-5515/50/i=6/a=064004} {\bibfield  {journal}
  {\bibinfo  {journal} {Nuclear Fusion}\ }\textbf {\bibinfo {volume} {50}},\
  \bibinfo {pages} {064004} (\bibinfo {year} {2010})}\BibitemShut {NoStop}%
\end{thebibliography}%

\end{document}